\documentclass[acmlarge,screen,9pt,nonacm]{acmart} 

\pdfoutput=1
\AtBeginDocument{%
  }

\usepackage{amsmath,amsfonts}
\usepackage{algorithmic}
\usepackage{array}
\usepackage{adjustbox}
\usepackage[caption=false,font=normalsize,labelfont=sf,textfont=sf]{subfig}
\usepackage{stfloats}
\usepackage{url}
\usepackage{verbatim}
\usepackage{graphicx}
\usepackage{booktabs}
\usepackage{multirow}
\usepackage{colortbl}
\usepackage{microtype}
\usepackage{afterpage}

\usepackage{textcomp}
\usepackage{xcolor}
\usepackage[ruled]{algorithm2e}
\usepackage{makecell}
\usepackage{tcolorbox}
\usepackage{rotating}
\usepackage{longtable}
\usepackage{bbding}
\usepackage{tabularx}
\usepackage{tcolorbox}
\usepackage{pifont}
\newcommand{\boxmargin}{1mm}
\newtcolorbox{myboxc}{
    colback=yellow!10!white, 
    colframe=gray!50,
    arc = 0pt, outer arc = 0pt,
    boxsep=0pt, left = 3pt, right = 0pt, top = 0pt, bottom = 0pt, 
    leftrule=3pt,
    bottomrule=0pt, toprule=0pt, rightrule=0pt,
    left = \boxmargin, right = \boxmargin, top = \boxmargin, bottom = \boxmargin
}
\usepackage{afterpage}

\newcommand{\newrevised}[1]{\bgroup\color{blue}#1\egroup}
\begin{document}

\title{Towards an Understanding of Context Utilization in Code Intelligence}




\author{Yanlin Wang}
\affiliation{
  \institution{Sun Yat-sen University}
  \city{Zhuhai}
  \country{China}
}
\email{wangylin36@mail.sysu.edu.cn}

\author{Kefeng Duan}
\affiliation{
  \institution{Sun Yat-sen University}
  \city{Zhuhai}
  \country{China}
}
\email{duankf@mail2.sysu.edu.cn}

\author{Dewu Zheng}
\affiliation{
  \institution{Sun Yat-sen University}
  \city{Zhuhai}
  \country{China}
}
\email{zhengdw5@mail2.sysu.edu.cn}

\author{Ensheng Shi}
\affiliation{
  \institution{Huawei Cloud Computing Technologies Co., Ltd.}
  \city{Beijing}
  \country{China}
}
\email{shiensheng@huawei.com}

\author{Fengji Zhang}
\affiliation{
  \institution{City University of Hong Kong}
  \city{Hong Kong}
  \country{China}
}
\email{fengji.zhang@my.cityu.edu.hk}

\author{Yanli Wang}
\affiliation{
  \institution{Sun Yat-sen University}
  \city{Zhuhai}
  \country{China}
}
\email{wangyli58@mail2.sysu.edu.cn}

\author{Jiachi Chen}
\authornote{Corresponding author.}
\affiliation{
  \institution{Sun Yat-sen University}
  \city{Zhuhai}
  \country{China}
}
\email{chenjch86@mail.sysu.edu.cn}

\author{Xilin Liu}
\affiliation{
  \institution{Huawei Cloud Computing Technologies Co., Ltd.}
  \city{Shenzhen}
  \country{China}
}
\email{liuxilin3@huawei.com}

\author{Yuchi Ma}
\affiliation{
  \institution{Huawei Cloud Computing Technologies Co., Ltd.}
  \city{Shenzhen}
  \country{China}
}
\email{mayuchi1@huawei.com}

\author{Hongyu Zhang}
\affiliation{
  \institution{Chongqing University}
  \city{Chongqing}
  \country{China}
}
\email{hyzhang@cqu.edu.cn}

\author{Qianxiang Wang}
\affiliation{
  \institution{Huawei Cloud Computing Technologies Co., Ltd.}
  \city{Shenzhen}
  \country{China}
}
\email{wangqianxiang@huawei.com}

\author{Zibin Zheng}
\affiliation{
  \institution{Sun Yat-sen University}
  \city{Zhuhai}
  \country{China}
}
\email{zhzibin@mail.sysu.edu.cn}

\renewcommand{\shortauthors}{Wang et al.}
\begin{abstract}
Code intelligence is an emerging domain in software engineering, aiming to improve the effectiveness and efficiency of various code-related tasks. Recent research suggests that incorporating contextual information beyond the basic original task inputs (i.e., source code) can substantially enhance model performance. Such contextual signals may be obtained directly or indirectly from sources such as API documentation or intermediate representations like abstract syntax trees can significantly improve the effectiveness of code intelligence.
Despite growing academic interest, there is a lack of systematic analysis of \textbf{context} in code intelligence. To address this gap, we conduct an extensive literature review of 146 relevant studies published between September 2007 and August 2024.
Our investigation yields four main contributions. (1) A quantitative analysis of the research landscape, including publication trends, venues, and the explored domains; (2) A novel taxonomy of context types used in code intelligence;
(3) A task-oriented analysis investigating context integration strategies across diverse code intelligence tasks; (4) A critical evaluation of evaluation methodologies for context-aware methods. 
Based on these findings, we identify fundamental challenges in context utilization in current code intelligence systems and propose a research roadmap that outlines key opportunities for future research.
\end{abstract}

\ccsdesc[500]{Software and its engineering~Automatic programming}

\keywords{Code intelligence, machine learning, deep learning, large language models, survey, context utilization}

\maketitle

\section{Introduction}
Code intelligence (CI) is a rapidly growing field that aims to harness the potential of artificial intelligence techniques, particularly deep learning, to extract and utilize knowledge embedded in large-scale code repositories\cite{xu2022survey,sun2024survey}. This domain supports critical software engineering tasks such as code generation~\cite{Athiwaratkun2023Multi}, program repair~\cite{N40}, and code summarization~\cite{U19}, all of which are crucial for improving software development, maintenance, and evolution~\cite{wang2024agents,zheng2025towards}. Recent advancements in the field have focused on incorporating contextual information, which provides additional information beyond the basic inputs required for these tasks. For example, CoCoGen~\cite{F4} leverages compiler feedback to iteratively localize project-level context and refine the generated code, ensuring that project-specific details are preserved. Similarly, Li et al.~\cite{N15} propose a context-based bug detection method that reduces false positives by integrating three types of context: program dependence graphs (PDGs), data flow graphs (DFGs), and abstract syntax trees (ASTs). Furthermore, RLPG~\cite{N30} takes the repository structure into account and trains a prompt proposal classifier to effectively utilize repository context for code completion, and RepoCoder~\cite{N31} proposes an iterative retrieval-generation pipeline to collect relevant code from the repository as additional context.

``context'' in code intelligence domain refers to additional information beyond the basic inputs required for code intelligence tasks, which can be sourced from code repositories and applied to enhance these tasks. \emph{Although recent advancements have shown promising results, to the best of our knowledge, no systematic survey exists on how context is modeled, integrated, and evaluated across code intelligence tasks.} This gap hinders the development of generalizable principles for context-aware systems. 
To address this, we conduct a systematic literature review of 146 studies on context utilization in code intelligence tasks between September 1, 2007, and August 31, 2024. We review seven code intelligence tasks, including four code-code tasks (clone detection, defect detection, code completion, program repair), two code-text tasks (code summarization, commit message generation), and a text-code task (code generation). We first map the research landscape, identifying key publication trends, prominent research venues, and the primary domains examined in these studies. Next, we propose a systematic categorization of the various types of context used, establishing a taxonomy to understand the current state of the art. We then analyze how these contexts are applied across different code-related tasks, providing insights into their practical applications and effectiveness. Finally, we assess the evaluation methodologies employed to measure the performance of context-aware code intelligence models. To guide our analysis, we investigate the following research questions (RQs):

\begin{itemize}
\item \textbf{RQ1: What are the trends and distributions of the studied publications?}
Our analysis of the 146 reviewed studies reveals a growing interest in utilizing context for code intelligence tasks, peaking in 2024. Of these studies, 52.7\% were published in conferences, while 23.3\% appeared in journals. Technique development dominates (83.6\%), while benchmarks (8.9\%), empirical studies (4.1\%), and case studies (3.4\%) remain underrepresented.

\item \textbf{RQ2: How are contexts categorized in code intelligence tasks?}
We investigate different types of context used in code intelligence and form a taxonomy. Based on whether contexts can be obtained directly, we categorize them into two major groups: direct context (e.g., source code, code diffs, API documents) and indirect context (e.g., ASTs, control flow graphs). Analysis results show that direct context is more frequently used than indirect context. Among the 14 identified forms of context, the most commonly used are \textit{source code}, \textit{abstract syntax trees (ASTs)}, and \textit{code diffs}. Although context is widely leveraged, there are still specific areas where the application is limited. The extent of context utilization varies significantly across tasks depending on how well the context aligns with the specific requirements of each task, revealing underexplored opportunities in context generalization across multiple tasks and the effective utilization of multiple context types within a single task.

\item \textbf{RQ3: What methods are used to preprocess and model context?} 
This RQ examines the methods that preprocess and model the context in code intelligence. For preprocessing, identifier splitting based on camel case conventions (25 studies) is the most common technique. Sequence-based deep learning models (49 studies) remain prevalent regarding context modeling. 
Additionally, the emergence of large language models (LLMs) emerge as a promising paradigm (31 studies).

\item \textbf{RQ4: How are context-aware code intelligence models evaluated?}
We assess the evaluation practices by examining the metrics, established benchmarks, and the open-source status for context-aware studies. For evaluation metrics, we document their definitions and the specific code intelligence domains in which they are applied. Notably, seven studies still rely on human evaluation, suggesting a need for more standardized metrics. In terms of benchmarks, we compile information on their corresponding tasks, programming languages, and how context is incorporated, identifying gaps in providing diverse context types across multiple programming languages. At last, we categorize the studies' open-source status into three dimensions: code, model weights, and datasets.

\end{itemize}

The main contributions of this study include:
\begin{itemize} 
    \item This is the first systematic review synthesizing context utilization in code intelligence tasks. 
    \item We build a taxonomy of context types and task-specific integration patterns. 
    \item We conduct quantitative analysis of methodological trends and evaluation practices. 
    \item We provide a research roadmap addressing scalability, generalizability, and assessment gaps, identifying key challenges and emerging opportunities for enhancing code intelligence through effective context utilization. 
\end{itemize}

Besides, we maintain a GitHub project\footnote{https://github.com/DeepSoftwareAnalytics/Repo-Context-Survey} associated with this survey, which contains the resources used to construct this paper. We will continue to update the resource, hoping that this approach will benefit the community and facilitate future research.

The remainder of this paper is organized as follows: Section~\ref{sec:background} introduces the background of code intelligence tasks. Section~\ref{sec:methodogy} describes the methodology used for paper selection. Sections~\ref{RQ1} to~\ref{RQ4} address RQs, respectively. Section~\ref{case_study} presents a case study that compares performance with the addition of context. Section~\ref{sec:relatedwork} presents related work and research focuses. Section~\ref{sec:challenges_and_opportunities} provides a detailed explanation of current challenges and potential opportunities. Section~\ref{sec:conclusion} concludes our study and outlines directions for future research in context utilization.

\section{Background} \label{sec:background}
Code intelligence, as defined by Wan et al.~\cite{Wan2024DL}, refers to the application of artificial intelligence techniques to extract knowledge from code repositories and assist programmers in various code-related tasks~\cite{shi2023towards,shi2022evaluation,zheng2024well,tao2021evaluation,wang2024repotransbench,wang2024beyond,gong2024cosqap,guo2024stop}. Code intelligence is pivotal in software engineering, equipping developers with advanced tools that reduce manual effort and boost productivity. Recent advances suggest that incorporating contextual information is the key to achieving better performance in code intelligence tasks. \textbf{Context} is the supplementary information beyond the \textbf{raw input} (the minimal, essential data a task requires), that a model leverages to enhance its comprehension and effectiveness. In this subsection, we briefly introduce seven key code intelligence tasks examined in this paper, including four code-code tasks (clone detection, defect detection, code completion, program repair), two code-text tasks (code summarization, commit message generation), and a text-code task (code generation). The specific definition and raw input for each task are detailed in Table~\ref{tab:tasks}.

\begin{itemize}
    \item \textbf{Code Generation}: Code generation, also known as program synthesis, involves automatically generating source code from natural language descriptions. Early approaches were limited to predefined scenarios~\cite{codegenUsage} or required manual template construction~\cite{codegenTemplates}, resulting in poor generalizability. The advent of machine learning~\cite{ling2016codegen,Yang2021codegen,rabinovich2017codegen} has enabled models to generate code for diverse domains, alleviating template dependency.  However, unlike human developers who intrinsically utilize context (e.g., compiler information, API documents), early ML models lacked mechanisms to incorporate such knowledge. Recent advances address this gap by integrating contextual cues. For example, API documentation~\cite{N6,N7} for semantic constraints, codebase analysis~\cite{F12,F15} for structural patterns, and compiler feedback~\cite{F4,F51} for syntactic validation. These context-aware techniques significantly improve code relevance and accuracy, demonstrating the potential of repository-grounded code generation.

    \item \textbf{Code Completion}: Code completion aims to generate syntactically correct and semantically coherent code fragments for unfinished code snippets. This task can be categorized along two orthogonal dimensions: (1) the type of uncompleted code, including subsequent code prediction~\cite{F29} and code infilling~\cite{codecmpfill}; (2) the granularity of completion, ranging from token-level~\cite{codecmpTokenLevel} and line-level~\cite{codecmpLineLevel} to function-level~\cite{N31} completion. Effective context utilization is pivotal across all types of code completion, as local variables, cross-file dependencies, and API calls necessitate a deep understanding of the development context. This requirement becomes particularly critical in repository-level completion~\cite{N31,F1,F2,F7}, where a retrieval module collects project-specific contexts to maintain global code consistency. Recent advances adopt two complementary strategies to enhance repository-level code completion: (1) Structural approaches like context graph construction~\cite{N29,F1} or code dependencies extraction to optimize retrieval methods; (2) Pipeline-centric methods employ iterative retrieval~\cite{N31}, dynamic judgement~\cite{F2,F17}, and reranking~\cite{F7} to search for efficient contexts and filter noisy context progressively. These innovations collectively advance the precision of context-aware code completion.

    \item \textbf{Code Summarization}: Code summarization, known as code comment generation, aims to automatically produce concise and informative descriptions based on given source code. This task significantly enhances software maintainability and team collaboration~\cite{zhang2022survey,wang2024sparsecoder}. Early approaches rely on rigid rule-based templates~\cite{codesummRuleBased}, and recent advances leverage deep learning to improve semantic understanding: CNNs extract local code features~\cite{codesummCNNBased}, RNNs model sequential logic~\cite{codesummRNNBased}, GNNs capture syntactic structures~\cite{F6}, and Transformers address long-range dependencies~\cite{F31}. However, even state-of-the-art architectures struggle to infer implicit knowledge (e.g., API usage conventions) from standalone code snippets. To bridge this gap, researchers enrich models with various types of context such as source code~\cite{N25} for syntactic patterns, API documents~\cite{N16} for semantic constraints, code comments~\cite{N59} for intent clarification, and UML diagrams~\cite{N35} for architectural context. By jointly analyzing repositories, context-aware models generate summaries that better articulate functional core and systemic roles, advancing precision and comprehensibility.

    \item \textbf{Commit Message Generation}: The commit message generation task takes code diffs as essential input and aims to automatically produce natural language summaries that concisely describe what code changes were made and why they were implemented~\cite{B3,tao2024KADEL}. High-quality commit messages are critical for software maintenance~\cite{zhang2024automatic}, as they document repository evolution and facilitate collaboration. Early rule-based methods~\cite{commitRule1} extract keywords from source code to compose messages but suffer from redundant text and fail to explain change rationales~\cite{commitgensurvey}. Approaches based on deep learning models~\cite{commitDL1,commitDL2,commitDL3}) improved fluency by learning from commit histories, yet often generate vague messages (e.g., "optimize performance") due to ignoring project-specific context. This context blindness causes style inconsistency and functional ambiguity. To address this, researchers first explore retrieving messages from similar historical commits~\cite{N45}, but retrieval alone struggles with novel code changes. Recent hybrid methods~\cite{N22,F42,F45} integrate neural generators with retrieval mechanisms: retrievers fetch context to guide style alignment, while generators synthesize commit messages with current changes and retrieved contexts. Such methods demonstrate promise in balancing specificity and consistency, advancing practical applicability.
    
    \item \textbf{Clone Detection}: Clone detection identifies duplicate or semantically equivalent code fragments within or across repositories. This capability supports critical applications like vulnerability propagation analysis~\cite{N10} and dataset deduplication~\cite{N31}. Early works focus on syntactic clone detection~\cite{clonedetectionsynt1,clonedetectionsynt2,clonedetectionsynt3}, which identifies code with structural similarity but fails to detect functional duplicates with differing implementations. As syntactic methods are mature, current researches shift toward semantic clone detection. Semantic clones exhibit equivalent functionality despite syntactic divergence, requiring models to reason beyond surface patterns. To bridge this gap, context-aware approaches integrate multi-modal signals. For example, execution traces~\cite{N11} provide dynamic behavior profiles to align code with varied control flows. Code comments~\cite{N44} encode developer intent, aiding in functional equivalence inference. API documentation~\cite{F41,F43} clarifies interface semantics, resolving abstraction-level ambiguities. By leveraging such context, models achieve higher precision in identifying semantically equivalent clones, even when syntactic similarity is absent.
    
    \item \textbf{Defect Detection}: Defect detection identifies code fragments that may introduce functional errors, security vulnerabilities, or performance degradation. Its predictive capability significantly enhances software reliability~\cite{thota2020survey}, driving active research across academia and industry. Current studies target diverse defect categories, such as just-in-time defects~\cite{justintimedefectsurvey}, reentry vulnerabilities~\cite{reentrydefect}, and security-critical flaws from the Common Vulnerabilities and Exposures (CVE)~\cite{F21}. While deep learning models~\cite{N15,N49,N56} and code similarity analysis~\cite{defectcodesim1,defectcodesim2} achieve notable progress, they still exhibit critical limitations without the guidance of context. For instance, without modeling cross-file dependencies in code property graphs, models may misclassify secure code as vulnerable when interacting with external libraries~\cite{N56}. To address this, recent works emphasize models with multi-level contextual information, such as bug reports~\cite{N51}, code change history~\cite{N63}), and context graph representations~\cite{N49,N56}. This enables more precise analysis in defect detection.

    \item \textbf{Program Repair}: Program repair task involves taking buggy source code as input and producing a corrected version while maintaining the original functionality. Existing approaches span rule-based methods using heuristic search~\cite{programrepairsearch1,programrepairsearch2}, manual restoration using templates~\cite{programrepairtemplates1,programrepairtemplates2}, and the application of semantic constraints~\cite{programrepairsemantic1,programrepairsemantic2}. The rise of deep learning also opens new research directions~\cite{N8,N40}, as deep neural networks can automatically extract features, potentially freeing researchers from labor-intensive rule writing and feature engineering. Automatic program repair remains challenging due to the complexity of handling interdependent code fragments and external libraries, which increases the need for contextual information to guide repairs. Incorporating contextual information from the source code~\cite{U01,U03,U08} helps narrow down potential bug sources, enabling more accurate identification and resolution of issues. Additionally, compiler information~\cite{N2} provides valuable context regarding syntax and execution, highlighting areas where code fails to compile or execute correctly. Despite the exciting progress in automatic program repair, a significant gap remains in achieving repository-level automatic program repair~\cite{F9}, which demands further exploration in context utilization and model improvement. 

\end{itemize}

\begin{table}[t]
\footnotesize
\centering
\begin{tabular}{|p{3.0cm}|p{8.8cm}|p{3.0cm}|}
\hline
\textbf{Task} & \textbf{Definition} & \textbf{Raw Input} \\
\hline
Code Generation & Generate corresponding source code based on natural language descriptions. & Natural language descriptions \\
\hline
Code Completion & Predict code elements (e.g., tokens, lines, or blocks) based on the current incomplete code. & Incomplete code units\\
\hline
Code Summarization & Generate concise and descriptive comments based on the given source code. & Source code \\
\hline
Program Repair & Automatically fix defects (bugs) in the given source code. & Buggy source code \\
\hline
Clone Detection & Identify code snippets with notable similarities. & Code snippet pairs \\
\hline
Defect Detection & Identify whether the given code snippet contains potential bugs or vulnerabilities. & Source code snippets \\
\hline
Commit Message Generation & Automatically generate descriptive commit messages for code changes (diffs). & Code diffs \\
\hline
\end{tabular}
\caption{the Definition and the Raw Input of Each Code Intelligence Task.}
\label{tab:tasks}
\end{table}

\section{Methodology}
\label{sec:methodogy}

To gain a comprehensive understanding of context utilization in the code intelligence domain, we follow previous work~\cite{sun2024survey} and select seven key tasks based on their input/output modalities. This structure ensures representative coverage across the three fundamental categories: four code-to-code tasks (clone detection, defect detection, code completion, and program repair), two code-to-text tasks (code summarization and commit message generation), and one text-to-code task (code generation).

\begin{figure*}[t]
    \centering 
    \includegraphics[width=0.8\linewidth]{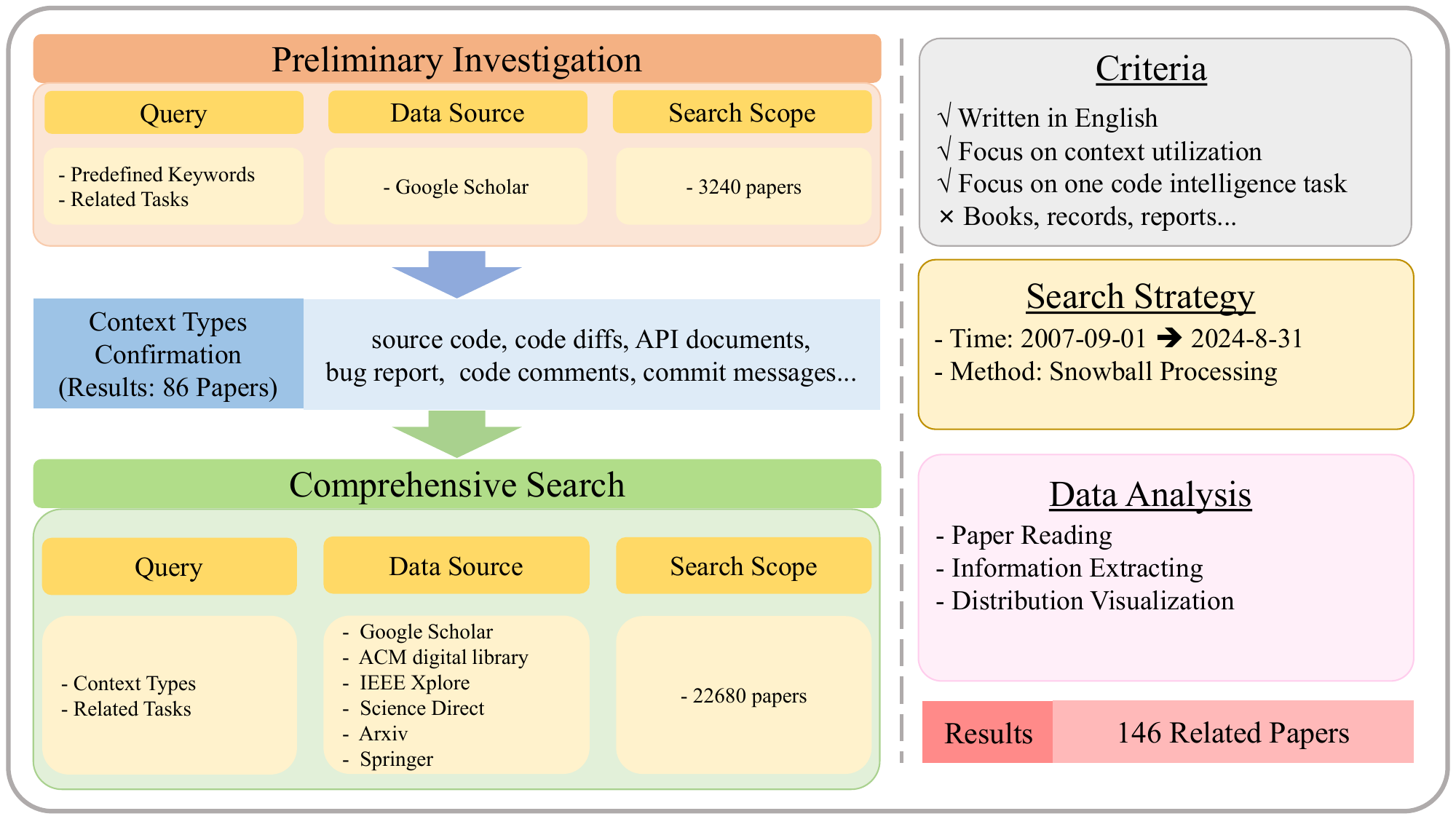}
    \Description[Short description]{The Pipeline in Collecting Relevant Papers}
    \caption{the Pipeline in Collecting Relevant Papers}
    \label{fig:Pip} 
\end{figure*}

Our systematic literature review employs a three-phase process methodology, as visually summarized in Figure~\ref{fig:Pip}. The process comprises (1) a preliminary investigation to establish the validity of context typology, (2) a comprehensive search for gathering context-relevant papers across seven code intelligence tasks, and (3) a systematic data analysis to address the defined research questions. 

This section elaborates on our methodological framework through four structured components. Section~\ref{sec:RQs} describes the research questions and their significance. Section~\ref{sec:LSS} outlines the literature collection process, focusing on the preliminary investigation and comprehensive search phases. Section~\ref{criteria} specifies our inclusion and exclusion criteria for study selection. Section~\ref{sec:analysis} presents the systematic approach for analyzing data extracted from the selected papers. 

\subsection{Research Questions} \label{sec:RQs}

To guide our systematic literature review and thoroughly investigate the evolution of context utilization in code intelligence tasks, we define the following research questions (RQs):

\begin{enumerate}
    \item RQ1: What are the trends and distributions of the publications studied?
    \item RQ2: How are contexts categorized in code intelligence tasks?
    \item RQ3: What methods are used to preprocess and model context?
    \item RQ4: How are context-aware code intelligence models evaluated?
\end{enumerate}

These research questions are designed to address key gaps in understanding which context is used and how it is applied across various code intelligence tasks. RQ1 examines historical trends and the distribution of publications, offering insights into the growing interest in context-aware models across seven key code intelligence tasks since 2007. RQ2 classifies context types and analyzes their impact across scenarios. RQ3 explores the preprocessing methods and techniques presented in recent studies to optimize context usage and analyzes the strengths and weaknesses of these approaches. Finally, RQ4 evaluates context-aware models, including the metrics and datasets used and their open-source status, to identify potential areas for improvement in model evaluation.

By answering these research questions, we aim to provide a comprehensive overview of context utilization in code intelligence and offer guidelines for selecting and applying techniques appropriate to specific contexts.

\subsection{Literature Search and Selection} \label{sec:LSS}

We follow the methodology illustrated in Figure~\ref{fig:Pip} to collect papers on using context in code intelligence. We employ a systematic, multi-phase process to define our search query and ensure completeness. Our search is divided into two primary stages: a preliminary investigation using initial keywords to broadly identify relevant work, followed by a comprehensive stage that uses context types gathered from the preliminary set to refine and complete the search. We identified 86 pertinent papers in the preliminary investigation and collected 163 papers after the comprehensive stage. After confirming their quality, we retained 146 papers.

\textbf{Search Query Definition.} Initially, we define two categories: Tasks (Specific, actionable research problems) and Keywords (Broad, high-level terms defining the research domain). To ensure representative coverage, the set of seven code intelligence tasks is pre-confirmed by structuring our selection based on three fundamental input/output modalities. To ensure maximal comprehensive search scope, we explicitly include aliases for code intelligence tasks (e.g., ``code summarization'' and ``code comment generation''; ``defect detection'' and ``vulnerability detection'')  as distinct entries in the tasks list. During the preliminary investigation, we employ an open card-sorting method~\cite{chen2021maintenance}, in which five authors leverage their domain expertise to generate keywords (broad conceptual terms, such as ``context'' and ``repository'') and consolidate the initial term pool. During the comprehensive stage, we categorize the identified context types after reviewing the preliminary papers. These context types replace keywords for further searches. The details of keywords, context types, and tasks are shown in Table~\ref{Search Type}.

Literature Retrieval and Completeness Assurance. Three rigorous methods are employed to ensure the completeness of our literature set.

\begin{itemize}
    \item \textbf{Phased Investigation:} Our literature search is executed in two distinct stages: a preliminary investigation and a comprehensive stage. Initially, the preliminary investigation searches 3,240 papers primarily using Google Scholar~\footnote{https://scholar.google.com/}. For the comprehensive stage, we expand the search data sources and utilize the context types identified in the preliminary investigation as refined keywords, resulting in a larger search scope of 22,680 papers.
    
    \item \textbf{Iterative Exhaustion:} In the preliminary investigation, we iteratively use the ``AND'' operator to combine the Keyword and Task for searching, until all combinations are explored. After reviewing the preliminary papers, we categorized the identified context types, as detailed in Table 1. In the comprehensive stage, we repeatedly paired each context type (acting as new refined keywords) with each task for further searches.
    
    \item \textbf{Two-Stage Snowballing:} For papers that meet our inclusion/exclusion criteria (outlined in Secton~\ref{criteria}), we apply a snowballing strategy~\cite{zheng2024LLM} in both investigation stages (reviewing both the references cited by this paper and the papers that cite this paper).
\end{itemize}

\begin{table}[t]
\centering
\setlength{\LTpre}{0pt}
\setlength{\LTpost}{0pt}
\begin{tabular}{lcp{12cm}}
\toprule

\textbf{Types}  & \textbf{Total}  & \textbf{Words}\\
\midrule
Keywords & 6&context, repository, API documents, project, file, UML\\
\midrule
\multirow{2}{*}{Context Types}&\multirow{2}{*}{14}&source code, code diffs, API documents, bug report, code comments, commit messages, AST, CFG, DFG, PDG, CPG, UML, compilation information, IDE \\
\midrule
\multirow{2}{*}{Tasks}&\multirow{2}{*}{9}&program repair, code generation, code completion, code summarization, code comment generation, clone detection, defect detection, vulnerability detection, commit message generation
\\ 
\bottomrule
\end{tabular}
\caption{Words Used for Searching}
\label{Search Type}
\Description[Short description]{Words Used for Searching}
\end{table}

\subsection{Inclusion and Exclusion Criteria} \label{criteria}
After identifying studies that align with our research objectives, it is necessary to filter out unqualified studies. To achieve this, we apply inclusion and exclusion criteria to assess the quality of candidate studies. Each retained research must satisfy all inclusion and exclusion criteria. This evaluation ensures that each retained research primarily focuses on using context to address one type of code intelligence task. The inclusion and exclusion criteria are as follows.

\Checkmark The paper must be written in English.

\Checkmark The paper must primarily focus on context utilization.

\Checkmark The paper must primarily focus on one task in code intelligence.

\XSolid Books, keynote records, reports, Master or Ph.D. theses, and grey literature are excluded.
 
\subsection{Data Extraction and Collection} \label{sec:analysis}

After filtering out irrelevant and duplicate papers, as shown in Figure~\ref{fig:Pip}, we extract relevant data and analyze it to address our four research questions. For RQ1, we record the year, type, publication venue, primary contributions, and task focus of each study. These details are vital for mapping the evolution of the research landscape, revealing historical trends, and understanding how various publication venues have influenced the development of context-aware code intelligence. For RQ2, we examine the types of context used and analyze trends in their application across seven code intelligence tasks. Collecting these data points is instrumental in identifying the diversity of context usage and in assessing its potential application scenarios. For RQ3, we review context preprocessing methods and context-aware models introduced in papers that present new techniques. This data collection sheds light on the methodological innovations and technical challenges involved in effectively leveraging context, which is crucial for advancing model development. For RQ4, we collate evaluation metric(s), dataset(s) created by the paper (excluding datasets created from external sources), and open-source status (availability of code, model weights, and datasets created by the paper).

cross-verification process for our collected data. The set of 146 papers was divided between two researchers (the paper's authors), who independently read their assigned papers and populated the extraction schema. Upon completion, the two researchers swapped their entire completed sets. Each researcher then performed a full verification of the other's work, checking every extracted data point against the original source paper. Any discrepancies were flagged and resolved through discussion and joint re-examination of the source paper by the two researchers. A third, senior author was designated as the arbiter to resolve any remaining disagreements, ensuring consistency across the entire dataset.

\section{RQ1: What Are the Trends and Distributions of the Publications Studied?}
\label{RQ1}

To understand the development trends of papers that utilize context in code intelligence, we analyze the trends and distributions of the reviewed publications. Our analysis focuses on three key aspects: publication dates, venues, and main contributions of the primary studies.

\subsection{Trends of Publication Year}
\begin{figure}[t]
    \centering 
    \includegraphics[width=0.8\linewidth]{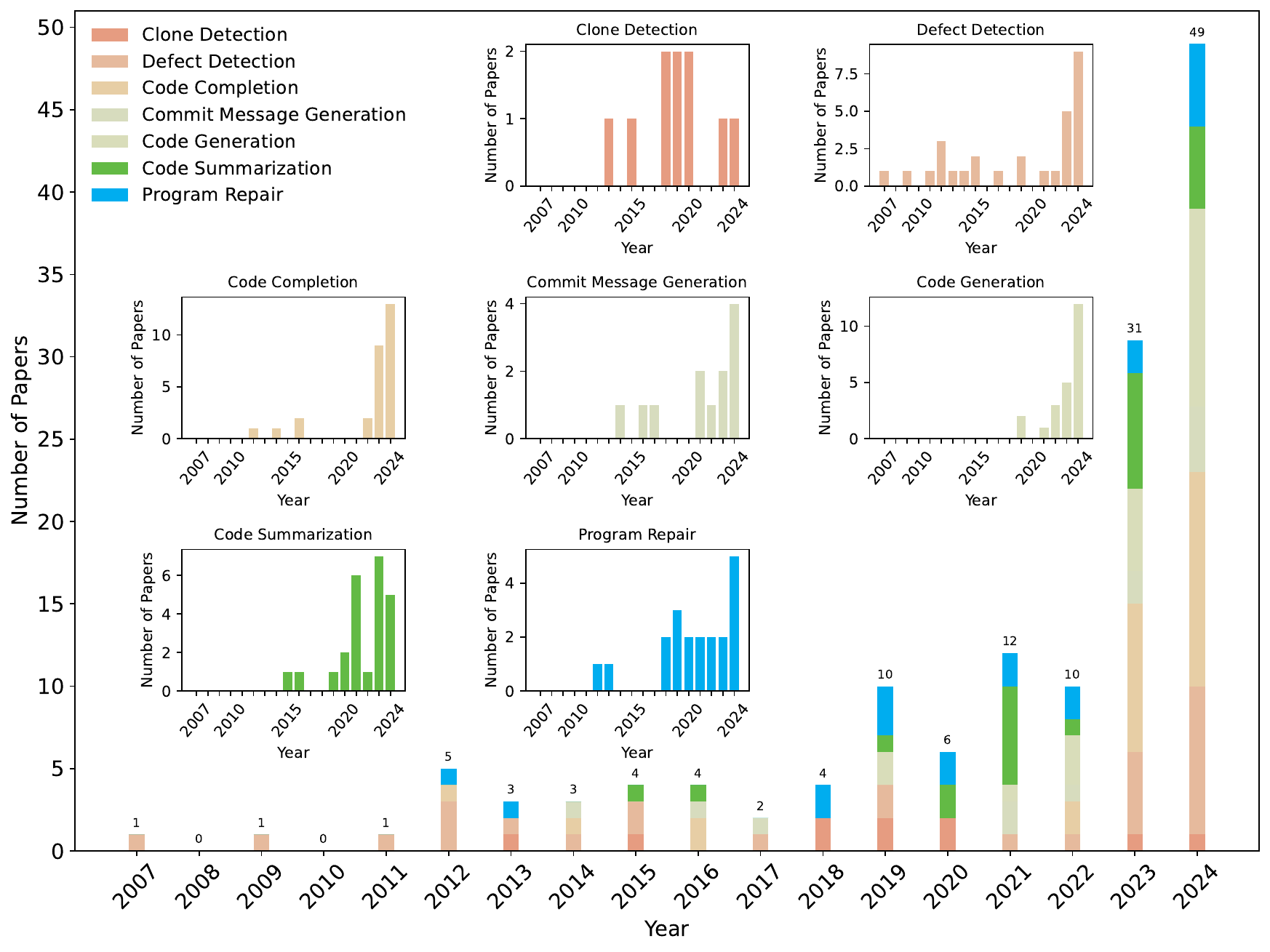}
    \Description[Short description]{Number of Papers Published Per Year}
    \caption{Number of Papers Published Per Year}
    \label{fig:YearPaper}

\end{figure}

Although the concept of ``context'' had been explored in natural language processing tasks before 2007~\cite{B2}, to the best of our knowledge, we did not find any papers prior to 2007 that used context to address code intelligence tasks in our investigation. Therefore, our analysis focuses on relevant papers published from 2007 to 2024. Figure ~\ref{fig:YearPaper} illustrates the publication trend of works employing context over the past eighteen years. To provide a clearer view of the trend in context utilization within code intelligence, we divide the research of these papers presented in these papers into three distinct phases.

The first phase, from 2007 to 2018, marks the early exploration of how code intelligence tasks process context. During this stage, researchers published fewer than five papers yearly, focusing mainly on the defect detection task. All other code intelligence tasks attract researchers to incorporate context into their models but with fewer or no published papers in seven code generation tasks. 

The second phase is from 2019 to 2022, during which more relevant work emerges. The average number of published papers per year rises to 9.5. This increase is related to earlier work that laid the foundation for context utilization and developing deep learning models, such as the transformer architecture~\cite{transformer}. At this stage, all seven types of code intelligence tasks begin to investigate how to fully utilize contextual information.

The third phase is from 2023 to 2024. In this phase, ChatGPT~\cite{chatgpt} initiates a wave of large language models, whose outstanding performance drives code intelligence tasks to further explore the use of context in large language models. During this phase, the number of papers increases substantially, with 31 papers published in 2023 and 49 in 2024 (as of August). Except for the clone detection task, all tasks show substantial growth in the number of papers. We suspect that clone detection may be less dependent on large language models.

\begin{figure}[t]
    \centering
    \includegraphics[width=0.8\linewidth]{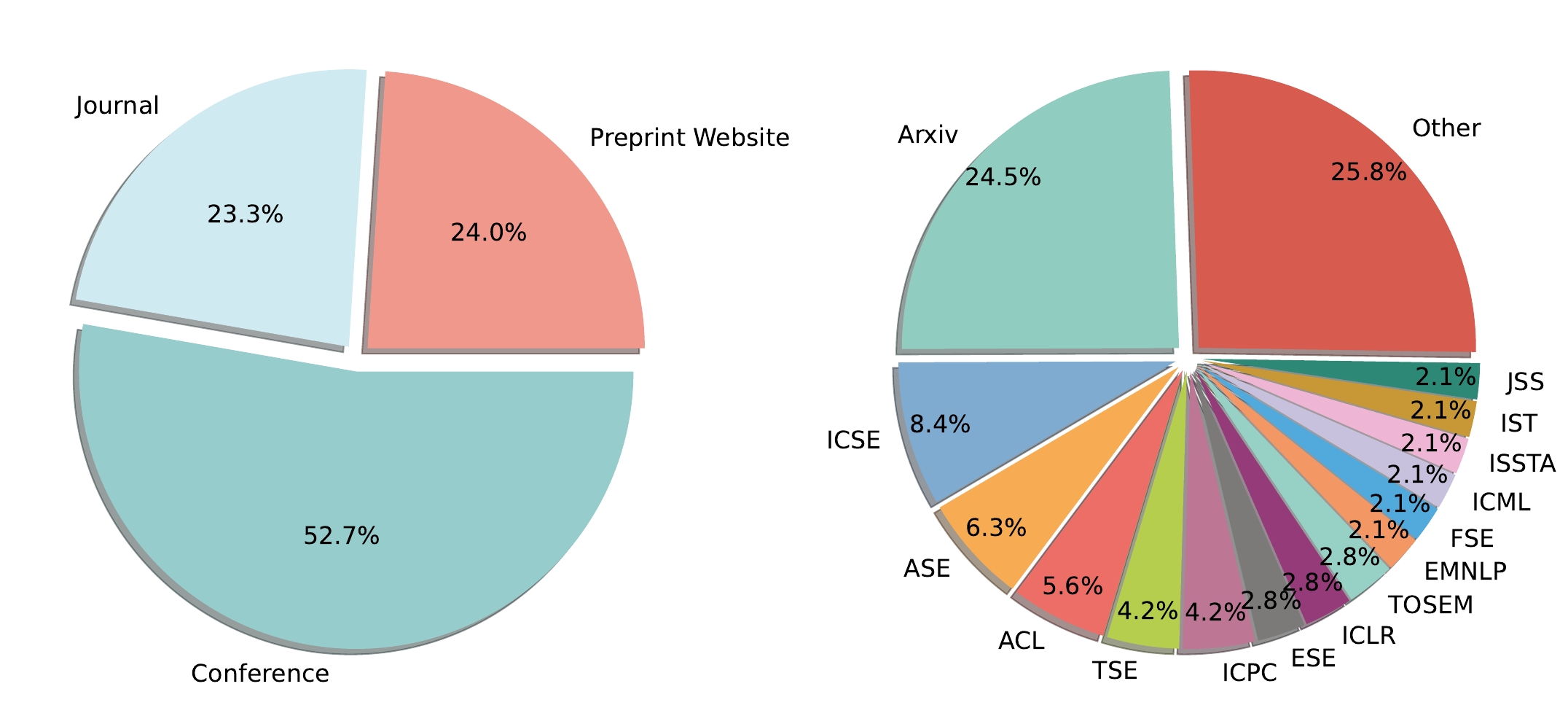}
    \Description[Short description]{(Left) Number of Publications Per Type. (Right) Number of Publications Per Venue.}
    \caption{(Left) Number of Publications Per Type. (Right) Number of Publications Per Venue.}
    \label{fig:Publication}
\end{figure}

\subsection{Trends of Publication Venues}

After reviewing 146 relevant papers,  we investigate the distribution across different publication types. As Figure ~\ref{fig:Publication} shows, the majority (52.7\%) are conference papers. Journal articles account for 23.3\%, representing less than a quarter of the total. The remaining 24.0\% are preprints, likely because many papers published in 2024 have not yet completed the journal or conference review process.

Furthermore, as Figure \ref{fig:Publication} shows, we explore the proportion of papers in each type of journal, conference, and preprint website. Several well-established conferences and journals in software engineering, such as ICSE (8.4\%), ASE (6.4\%), TSE (4.2\%), and ICPC (4.2\%), hold a significant proportion of the papers. Meanwhile, ACL (5.6\%), a prestigious conference in natural language processing, ranks third in paper count among all journals and conferences. Arxiv, a preprint website, maintains a leading share with 24.8\%. Journals or conferences that include no more than two papers are categorized as ``Other'', which accounts for the most significant proportion, over one-quarter, indicating that relevant papers are dispersed across various journals or conferences.

\subsection{Types of Main Contributions of the Studied Papers}

\begin{table}[t]
	\centering
        \caption{The Definition of Four Main Contributions in Primary Studies.}
	\begin{tabularx}{\textwidth}{lX}
		\toprule
		\textbf{Main contribution} & \textbf{Definition} \\
		\midrule
		\textbf{New Technique} & The study proposes a viable solution or devises an innovative framework to tackle specific issues.
		\\
		\midrule
		\textbf{Empirical Study}& The research gathers primary data and conducts quantitative and qualitative analyses to investigate intriguing discoveries.
        \\
        \midrule
        \textbf{Case Study}& The research analyzes issues by examining one or more distinct cases.
        \\
        \midrule
        \textbf{Benchmark}& The research aims to establish reference datasets, propose evaluation mechanisms, and identify leading models for comparison, with the goal of pinpointing areas for improvement and advancing the field's understanding and capabilities.
        \\
		\bottomrule
        \label{tab:ContributionDefinition}
	\end{tabularx}
\end{table}

\begin{figure}[t]
    \centering
    \includegraphics[width=0.8\linewidth]{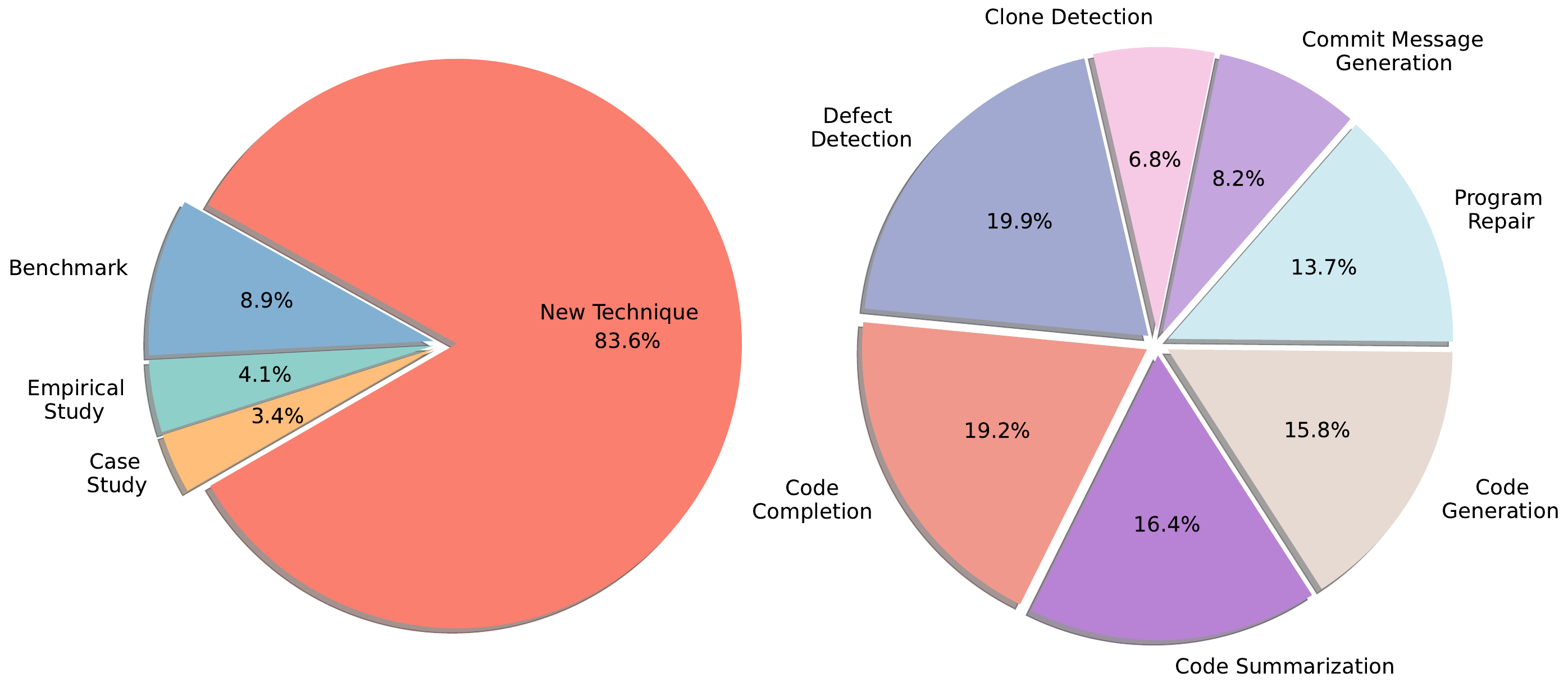}
    \Description[Short description]{(Left) Number of Papers Per Type of Main Contribution. (Right) Number of Papers Per Tasks.}
    \caption{(Left) Number of Papers Per Type of Main Contribution. (Right) Number of Papers Per Tasks.}
    \label{fig:CombinedTypeTaskPaper}
\end{figure}

To the best of our knowledge, we have not found any survey about context in 7 code intelligence tasks. Summarizing the main contribution of each paper, we categorize these studies into four types, i.e., new technique, Benchmark, empirical study, and case study. The definition of each main contribution is listed in Table ~\ref{tab:ContributionDefinition}. After analyzing all papers, as shown in Figure ~\ref{fig:CombinedTypeTaskPaper}, new technique papers constitute the majority, accounting for 83.6\% in total. 8.9\% is the benchmark section, followed by empirical studies, which occupy 4.1\%. Only 3.4\% is the case study, which is the least represented.

We simultaneously analyze the proportions of each type of code intelligence task, as shown in Figure~\ref{fig:CombinedTypeTaskPaper}. Top-3 code intelligence tasks are defect detection (19.9\%), code completion (19.2\%), and code summarization (16.4\%), while code generation occupies 15.8\% and program repair contains 13.7\%. Commit message generation and clone detection take the minor parts, 8.2\%, and 6.8\%, respectively.

\begin{center}
    \begin{myboxc}{\textbf{\ding{43} RQ1 Summary: }
    \begin{enumerate} 
        \item The survey results show a steady increase in the number of context-relevant papers across the seven code intelligence tasks, with a peak in 2024.
        \item More than half of the primary studies are published at conferences, while nearly half remain on preprint platforms. These publications are predominantly found in software engineering venues, with notable contributions in natural language processing conferences such as ACL.
        \item Most of the papers focus on proposing new techniques, followed by creating benchmarks and conducting empirical studies. The distribution of papers across the seven code intelligence tasks is relatively balanced.
    \end{enumerate}
    }
    \end{myboxc}
\end{center}

\section{RQ2: How Are Contexts Categorized in Code Intelligence Tasks?}
\label{RQ2}

To categorize context in code intelligence, we divide it into two types based on how it is obtained: direct and indirect. Direct context refers to information that can be accessed without further processing, such as source code, API documentation, and commit messages. Indirect context, however, requires additional processing or transformation, examples of which include abstract syntax trees (ASTs), control flow graphs (CFGs), and compiler-generated information. This section will explore the different context categories and examine trends in their use in code intelligence tasks.

\subsection{Types of Context} \label{Types of Context}

\begin{figure}[t]
    \centering 
    \includegraphics[width=0.8\linewidth]{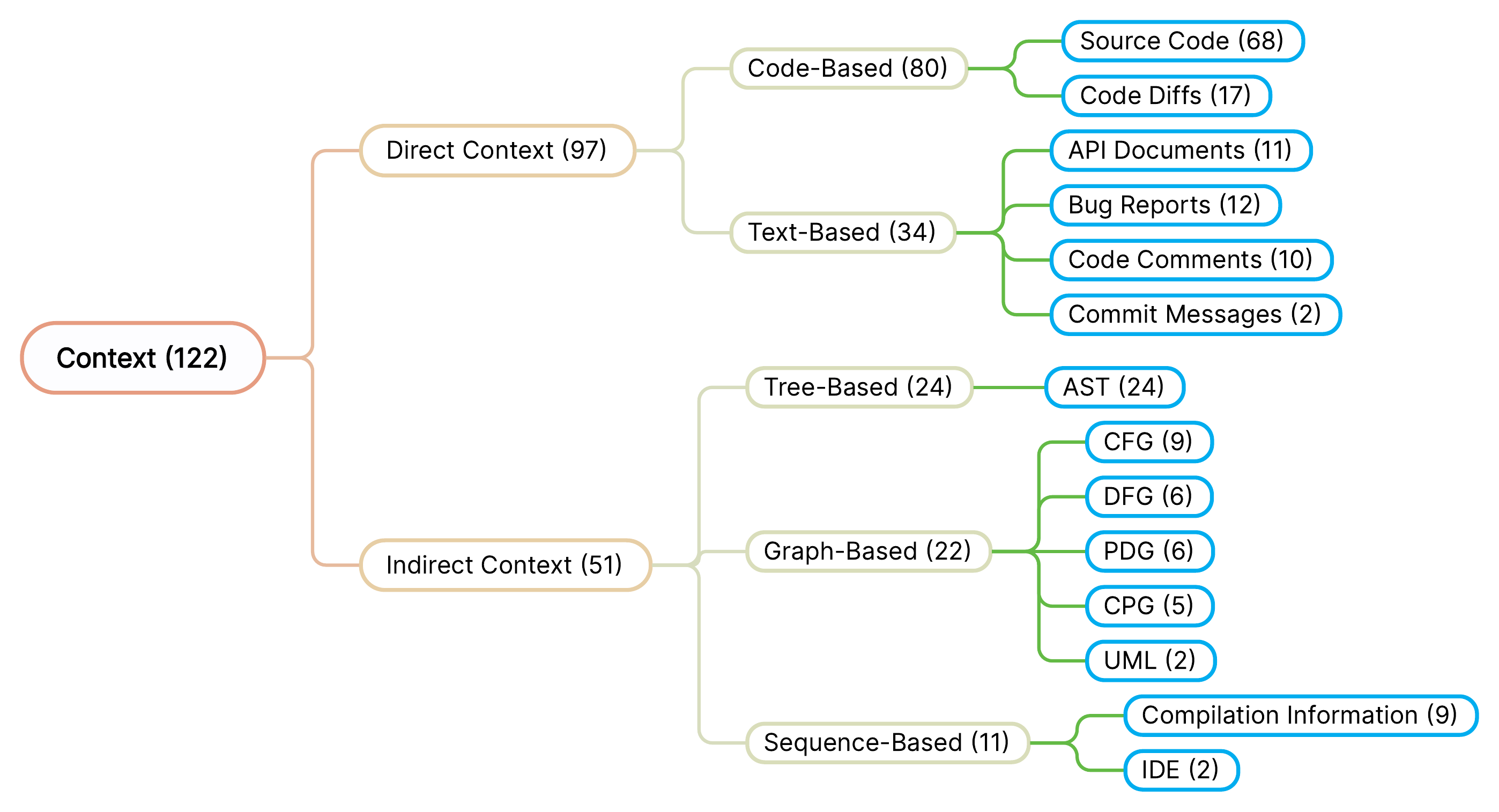} 
    \Description[Short description]{Types of Context Used in the Seven Studied Code Intelligence Tasks}
    \caption{Types of Context Used in the Seven Studied Code Intelligence Tasks}
    \label{fig:ContextTree} 
\end{figure}

As shown in Figure ~\ref{fig:ContextTree}, we summarize all types of contexts into a tree diagram. The number after the context name indicates how many new technique papers use this context, while benchmark, case study, and empirical papers are excluded. The combined count of direct context utilization (97 papers) and indirect context utilization (51 papers) exceeds the total number of papers (122 papers) because 26 papers utilize both direct and indirect contexts. Overall, direct contexts are more prevalent than indirect contexts. However, researchers discovered that transforming direct context into various indirect contexts in specific scenarios can improve model performance and efficiency. For example, Zhang et al.~\cite{N49} propose a graph neural network based on control flow graphs to detect condition-related bugs, which significantly outperforms state-of-the-art methods. Li et al.~\cite{N15} embed abstract syntax trees, data flow graphs, and program dependency graphs to capture local and global context for bug detection. These representations outperform the other representations by up to 206\% in accuracy.

\begin{figure}[t]
    \centering
    \includegraphics[width=0.85\linewidth]{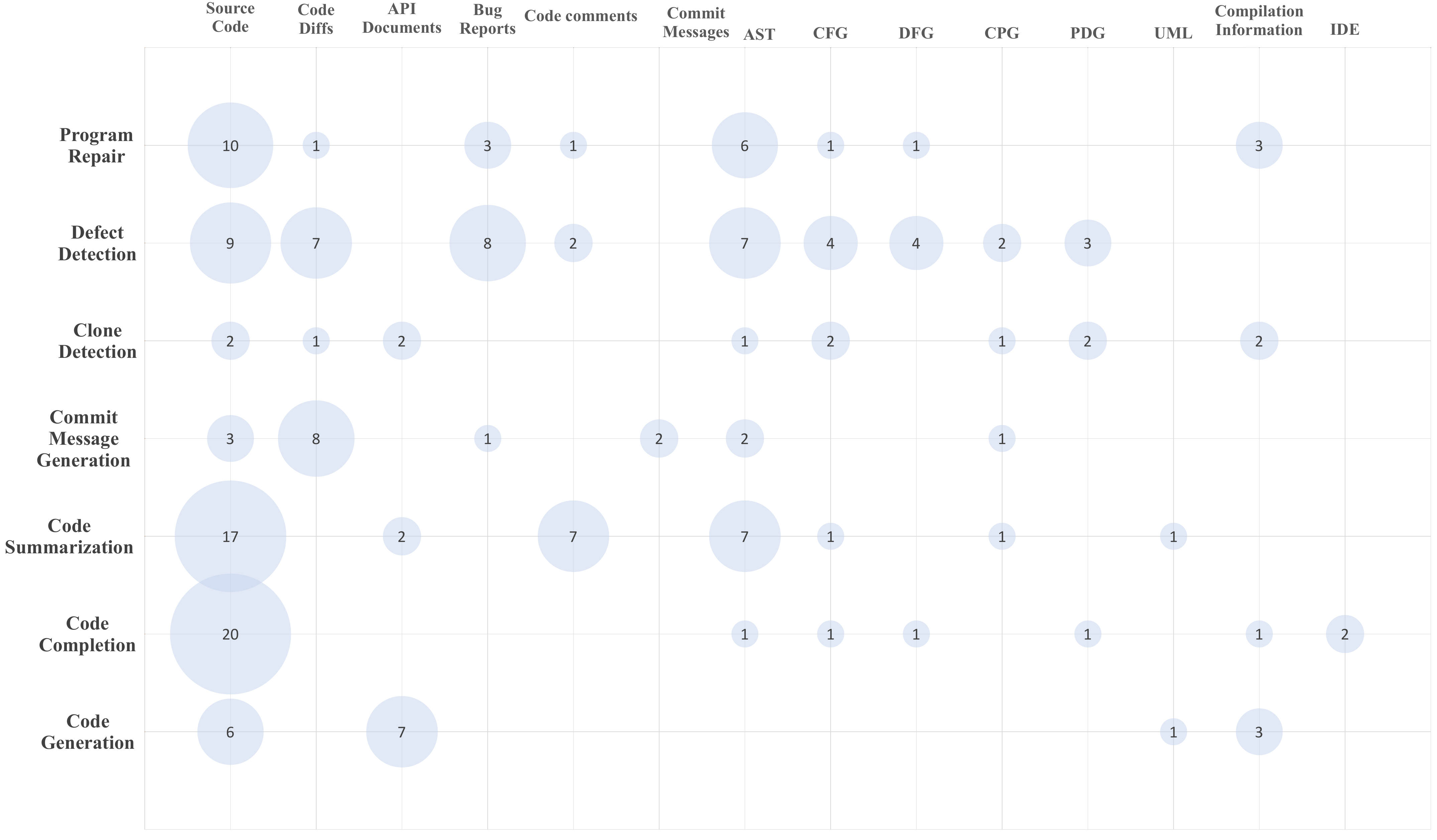}
    \Description[Short description]{Number of Context Utilized in Each Code Intelligence Task}
    \caption{Number of Context Utilized in Each Code Intelligence Task}
    \label{fig:ContextBubble} 
\end{figure}

\subsubsection{Direct Context}

Direct context is composed of code-based and text-based elements. The former mainly includes source code and code diffs, while the latter encompasses API documents, bug reports, code comments, and commit messages.

Eighty papers adapt code-based contexts, which are the most numerous of all forms of contexts, including sixty-eight papers for source code and seventeen papers for code diffs. Source code, the most popular form of context, constructs the basis for program execution. Mining and extracting source code context mitigates hallucination and provides more accurate answers by offering models with specific domain knowledge in repositories~\cite{F10,N31}. Code diff context presents changes in source code over time or across versions, providing more fine-grained information than source code context in tasks such as defect detection~\cite{N63} and commit message generation~\cite{F42}. 

Text-based contexts provide natural language descriptions of code information, including API documents (11 papers), bug reports (12 papers), code comments (10 papers), and commit messages (2 papers). API documents provide detailed API information; bug reports describe issues found in source code; code comments explain specific parts of code; and commit messages summarize changes made to repositories, all showing one or several dimensions of information in repositories.

\subsubsection{Indirect Context}
Indirect context can be categorized into three types: tree-based, graph-based, and sequence-based contexts. Tree-based context includes AST (24 papers), and constructing ASTs showcases the structure of code. The graph-based context includes CFG (9 papers), DFG (6 papers), PDG (6 papers), CPG (5 papers), and UML (2 papers), providing important information about source code, such as data flow, control flow, logic, and structure. Sequence-based context includes compilation information (9 papers) and IDE information (2 papers). Compilation information offers insight into the dynamics of code execution, and IDE information provides information such as programmers' usage patterns.

\subsection{Trends of Context} \label{Trends of Context}

Figure ~\ref{fig:ContextBubble} illustrates a detailed investigation of context applications in 7 code intelligence tasks. 
In figure~\ref{fig:ContextBubble}, the x-axis represents different code intelligence tasks, and the y-axis indicates various context types. The numbers represent the number of papers using that type of context for the corresponding task. The larger the bubble, the greater the number of papers, suggesting that the corresponding context type is more prevalent for the task. 
Although context is frequently leveraged, there are specific areas where its application is either insufficient or remains unexplored. Context selection is often closely linked to the particular code intelligence task. For instance, bug reports are widely utilized in program repair~\cite{N21,N39,N60} and defect detection~\cite{N13,N50,N53} but are less prevalent in other types of code intelligence tasks such as code generation and code completion.

Besides, we find that source code is the most popular form of context, utilized by all seven code intelligence tasks, followed by AST (utilized by six tasks) and CFG (utilized by five tasks). In all 13 kinds of context, defect detection employs the widest variety of context (9 types), while program repair and clone detection both select eight kinds of context, respectively. The variation in the amount of context utilized across different code intelligence tasks is due to how well the context information aligns with the specific requirements of each task.

However, some kinds of context are less common. Commit message and IDE information are utilized twice by commit message generation~\cite{N45,F44} and code completion~\cite{F18,F30}, respectively. UML is selected by code generation~\cite{N19} and code summarization~\cite{N35}. The limited application of these contexts is largely attributable to the scarcity of publicly available, standardized datasets and the inherent difficulty in translating their information for specific tasks. Besides, the reasons, such as data privacy for IDE logs and the high abstraction level of UML diagrams, also limit the usage of this context.

\begin{center}
    \begin{myboxc}{\textbf{\ding{43} RQ2 Summary: }
    \begin{enumerate}
        \item Among the 14 identified forms of context, source code, abstract syntax trees (ASTs), and code diffs are the three most commonly used. Researchers tend to favor direct context over indirect context.
        
        \item Different code intelligence tasks utilize a varying range of context, depending on how effectively the context meets the specific needs and requirements of each task. 
        
        \item Although many studies have explored the use of context in code intelligence tasks, certain areas that utilize specific types of context remain underexplored.
    \end{enumerate}
    }
    \end{myboxc}
\end{center}

\section{RQ3: What Methods Are Used to Preprocess and Model Context?}

\begin{figure}[t]
    \centering 
    \includegraphics[width=0.8\linewidth]{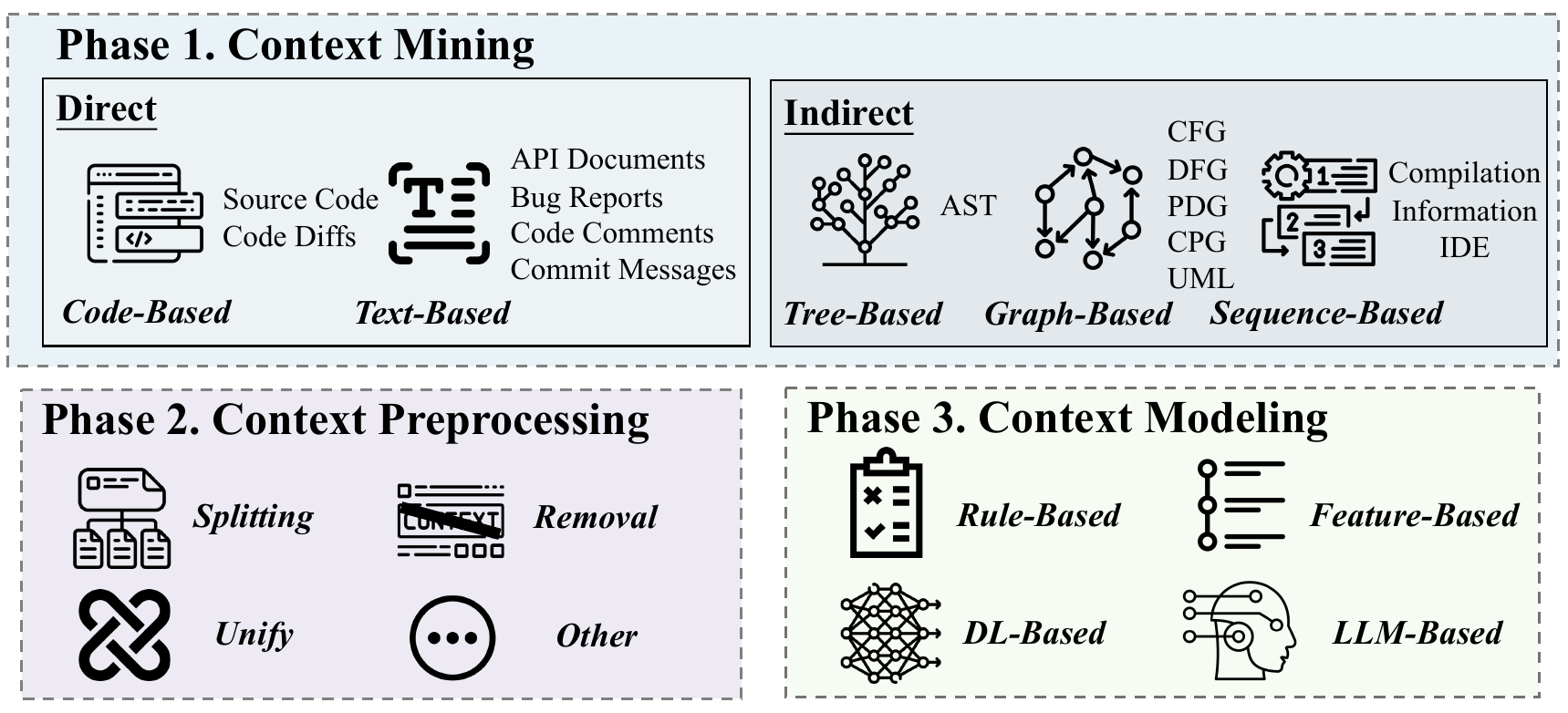}
    \Description[Short description]{The Pipeline of Context Utilization}
    \caption{The Pipeline of Context Utilization}
    \label{fig:End2End} 
\end{figure}

While context provides valuable information, it must be carefully processed and modeled to be effectively used in code intelligence tasks. Preprocessing refers to the techniques applied to the raw context data to structure and transform it into a usable input representation. Modeling involves the architectural workflow and methods used to integrate this already-preprocessed context representation into models to inform and improve task performance. This section examines how researchers process context, focusing on both preprocessing methods and modeling strategies used in various code intelligence tasks. {Figure~\ref{fig:End2End} illustrates the end-to-end pipeline of context utilization across code intelligence tasks, including context mining, context preprocessing, and context modeling.

\subsection{Context Preprocess}

To intuitively understand the effect of context preprocessing in each paper, we categorize the preprocessing operators into four types: splitting, irrelevance removal, unifying, and other operators. The details of the preprocessing steps are shown in Table ~\ref{tab:dataPre}.

\subsubsection{Splitting}

Splitting operators aim at segmenting raw data into token sequences that can be effectively recognized by models. These operators are designed to avoid generating an excessively large vocabulary containing numerous low-frequency words. Besides, splitting operators mitigates out-of-vocabulary issues, enabling models to better handle unseen inputs and enhancing their generalization capability. 

Byte pair encoding (BPE)~\cite{gage1994new} is a widely used splitting method in natural language processing. BPE reduces the number of tokens by iteratively converting frequently occurring data into new units, alleviating the training burden while maintaining data accuracy. For example, Liu et al.~\cite{N61} employ BPE method to segment both natural language and source code data, facilitating the generation of feature embeddings. This technique can be applied in code intelligence tasks to process code-relevant context, such as API documentation~\cite{N3,N4,N6}, compiler information~\cite{N1}, code comments~\cite{N61}, source code contexts~\cite{N29,N30,N31}.

Programming languages are highly structured and adhere to stricter syntax rules than natural language descriptions. To facilitate program comprehension, programmers often adopt CamelCase or SnakeCase as naming conventions for coding identifiers, such as variables, functions, and classes. These conventions offer practical guidelines for splitting identifiers: CamelCase identifiers can be divided by detecting transitions between lowercase and uppercase letters, while SnakeCase identifiers are split at underscores ('\_'). Both techniques are widely used and effective in preprocessing source code context~\cite{U01,N35,N37}. For example, Wang et al.~\cite{N35} employ the CamelCase rule to split class names in UML, while Bansal~\cite{N37} utilizes the CamelCase rule to split cross-file source code tokens.

Additionally, segmentation rules based on non-alphabetic~\cite{N7,N18,N58} or non-numeric symbols~\cite{N7,N33} can serve as supplementary approaches for segmenting both code and natural language content. These methods are widely used for their simplicity, as they do not rely on complex lexicons or language models and serve as an initial approach to token segmentation.

Another approach is to use a tokenizer from an established library or provided by language models. Examples of established libraries include NLTK  ~\footnote{https://github.com/nltk/nltk}~\cite{N48,N50}, Javalang ~\footnote{https://github.com/c2nes/javalang}~\cite{N49} and Pygments ~\footnote{https://github.com/pygments/pygments}~\cite{N42}. Examples of tokenizers provided by language models include those from StarCoder~\cite{F1,F16,F27}, GPT-4~\cite{F13}, and RoBERTa~\cite{F19}. A tokenizer segments continuous text into discrete tokens, assigns each token a unique index, and maps the index to token embeddings for further processing.

\begin{table}[t]
\centering
\footnotesize
\caption{Context Preprocess Methods in Seven Code Intelligence Tasks}
\renewcommand{\arraystretch}{1.3}
\setlength{\tabcolsep}{3pt}
\belowrulesep=0pt
\aboverulesep=0pt
\begin{tabular}{|c|c|p{2.2cm}|c|}
    \toprule
    \textbf{Operator} & \textbf{Sub-operator} & \textbf{Paper} & \textbf{\#Studies}  \\
    \midrule
    \multirow{6}{*}{Splitting}
    & BPE & \multicolumn{1}{p{9cm}|}{\cite{U07,N1,N3,N4,N5,N6,N29,N30,N31,N61,N65,F3}} & 12  \\
    \cline{2-4}
    & Camel\_Case & \multicolumn{1}{p{9cm}|}{\cite{U01,U02,U14,U19,N16,N24,N26,N30,N35,N37,N38,N42,N45,N48,N50,N51,N52,N53,N55,N57,N58,N59,N60,N63,F6} } & 25 \\
    \cline{2-4}
    & Snake\_Case & \multicolumn{1}{p{9cm}|}{\cite{U02,U19,N16,N24,N26,N30,N35,N37,N38,N42,F6}} & 11 \\
    \cline{2-4}
    & Base on Non-Alphabet Symbols & \multicolumn{1}{p{9cm}|}{\cite{U19,N7,N18,N33,N58}} & 5 \\
    \cline{2-4}
    & Base on Non-Numeric Symbols & \multicolumn{1}{p{9cm}|}{\cite{N7,N33} } & 2 \\
    \cline{2-4}
    & Base on Tokenizer & \multicolumn{1}{p{9cm}|}{\cite{N42,N48,N49,N50,F1,F13,F15,F16,F19,F27,F28}} & 11 \\
    \midrule
    \multirow{9}{*}{Removal}
    & Stopword Removal & \multicolumn{1}{p{9cm}|}{\cite{N51,N52,N53,N57,N60,N63} } & 6 \\
    \cline{2-4}
    & Logical Operators Removal & \multicolumn{1}{p{9cm}|}{\cite{N41} } & 1 \\
    \cline{2-4}
    & White Spaces Removal & \multicolumn{1}{p{9cm}|}{\cite{F6,N41}} & 1 \\
    \cline{2-4}
    & Punctuations Filtering & \multicolumn{1}{p{9cm}|}{\cite{N16,N24,N35,N38,N39,N41,N57,N60,N63,F6}} & 10 \\
    \cline{2-4}
    & Comment Removal & \multicolumn{1}{p{9cm}|}{\cite{N42,F31,F45}} & 1 \\
    \cline{2-4}
    & Empty Line Removal & \multicolumn{1}{p{9cm}|}{\cite{N42,F49}} & 2 \\
    \cline{2-4}
    & Non-Word Removal & \multicolumn{1}{p{9cm}|}{\cite{N38} } & 1 \\
    \cline{2-4}
    & Code Diff Removal & \multicolumn{1}{p{9cm}|}{\cite{F37}} & 1 \\
    \cline{2-4}
    & Symbol Removal & \multicolumn{1}{p{9cm}|}{\cite{F31}} & 1 \\
    \midrule
    \multirow{5}{*}{Unifying}
    & Lowercase & \multicolumn{1}{p{9cm}|}{\cite{N16,N24,N35,N37,N38,N60}} & 6 \\
    \cline{2-4}
    & Stemming & \multicolumn{1}{p{9cm}|}{\cite{N39,N50,N52,N53,N57,N60,N63}} & 7 \\
    \cline{2-4}
    & String Replacement & \multicolumn{1}{p{9cm}|}{\cite{U03,N18,N25,N35,N38,N42,N48}} & 7 \\
    \cline{2-4}
    & Number Replacement & \multicolumn{1}{p{9cm}|}{\cite{U03,N18,N35,N42,N48}} & 5 \\
    \cline{2-4}
    & Alpha Renaming & \multicolumn{1}{p{9cm}|}{\cite{N40}} & 1 \\
    \midrule
    \multirow{2}{*}{Other Operator} & \multirow{2}{*}{--} & \multicolumn{1}{p{9cm}|}{\cite{U04,U05,U08,U11,N17,N20,N22,N23,F1,F4,F8,F10,F11,F12,F13,F16,F18,F19,F20,F22,F23,F24,F26,F27,F29,F30,F38,F39,F40,F41,F43,F47,F48,F52,F57}} & \multirow{2}{*}{35} \\
    \bottomrule 
\end{tabular} 
\label{tab:dataPre}
\end{table}

\subsubsection{Removal}

To filter noise in data, researchers commonly remove irrelevant information to enhance model performance. 

For natural language data, stopwords are often removed due to their high frequency and little meaningful contribution (e.g., "a" and "the")~\cite{N51,N52,N53,N57,N60,N63}. For code-formatted data, heuristic rules can be designed to clean the data effectively. Irrelevant elements, such as logical operators~\cite{N41}, trailing white spaces~\cite{N41}, punctuation marks~\cite{N16,N24,N35,N38,N39,N41,N57,N60,N63}, comments~\cite{N42}, empty lines~\cite{N42}, and non-word characters~\cite{N38}, can also be removed to simplify the data and reduce noise. 

These removal steps are crucial for ensuring that models focus on meaningful patterns rather than irrelevant details, ultimately improving their overall performance.

\subsubsection{Unifying}

The unifying operator can be used to normalize data in the preprocessing step, helping models to organize data and handle sensitive information, thereby enhancing the model's generalization ability. 

To organize natural language formatted data, common text unifying methods include using lowercase~\cite{N16, N24, N35} and applying stemming~\cite{N39,N50,N52,N53,N57,N60,N63}. These works simplify the data by standardizing variations in word form. Lowercasing ensures that all tokens are converted to lowercase while stemming groups of words with the same core meaning by retaining the root form of words. Both methods help models focus on the core meaning of the data rather than extra information.

To organize code-based data, alpha renaming~\cite{N40} is an effective method for unifying data. Alpha renaming ensures that variable names are unique, preventing ambiguity when evaluating expressions or applying transformations. This operation helps avoid naming conflicts and improves the clarity of source code.

In addition to organizing data, handling sensitive information is crucial to ensure data purity and model security. Researchers replace numbers and strings in source code with unique tokens such as <num>~\cite{U03,N18,N35,N42,N48} and <str>~\cite{U03,N18,N25,N35,N38,N42,N48}. Numbers and strings can vary in length and format, requiring additional handling methods. Masking specific values encourages models to focus on the data type. Additionally, replacing them with unique tokens helps protect private data, since numbers and strings may sometimes contain sensitive information. 

\subsubsection{Other}

Additionally, 35 papers employ alternative preprocessing methods, reflecting the complexity of context preprocessing in code intelligence tasks.

One of the preprocessing methods involves using ASTs to extract information from source code. In this case, ASTs are not considered part of the context but are used to extract semantic code knowledge from repositories~\cite{F4,F12}. This narrows the scope of source code context while facilitating retrieval by constructing semantic graphs~\cite{F1}.

Besides, researchers also use static analysis to convert source code into XML format~\cite{F52} or apply program slicing~\cite{F57}, enabling further analysis and processing. In addition, manual expertise~\cite{F47} and rule-based methods~\cite{F38} are also applied in the preprocessing step to localize context and extract information.

These diverse preprocessing techniques highlight the intricate and varied approaches required to handle the context in code intelligence tasks, emphasizing the need for tailored strategies depending on the specific task and data characteristics.

However, 58 works do not clearly document their context preprocessing steps. This lack of clarity may hinder the accurate reproduction of results, thereby reducing the reliability of these works. We strongly urge future research to clearly and thoroughly document preprocessing steps to ensure transparency and reproducibility.

\subsection{Context Modeling Methods in Seven Code Intelligence Tasks}

\begin{table} [t]
\renewcommand{\arraystretch}{1.3}
\belowrulesep=0pt
\aboverulesep=0pt
\newcommand{\tabincell}[2]{\begin{tabular}{@{}#1@{}}#2\end{tabular}}
\centering 	
\footnotesize
\caption{Context Modeling in Seven Code Intelligence Tasks}
\label{tab:input forms in datasets}

\begin{tabular} {|c|c|c|>{\centering\arraybackslash}p{7cm}|c|c|}
\toprule
\textbf{Family} & \textbf{Sub-Family} & \textbf{Model Name} & \textbf{Paper} & \textbf{\#Studies} & \textbf{Total} \\
\hline
\multirow{2}{*}{Rule-based} &\multirow{2}{*}{--}
&\multirow{2}{*}{--} & \multicolumn{1}{p{7cm}|}{\cite{N10,N11,N12,N17,N19,N39,N43,N45,N46,N52,N55,N57,N59,N60,N63,U05,U06,U08,U09,U12,U13,U14,U15,U16,U17,U18,F24,F43,F55}} &\multirow{2}{*}{29} &\multirow{2}{*}{29} \\
\midrule

\multirow{4}{*}{Feature-based} & \multirow{4}{*}{--}
& SVM &\multicolumn{1}{p{7cm}|}{\cite{N21,N51,N53}} & 3 & \\ 
\cline{3-5}
&& Decision Tree&\multicolumn{1}{p{7cm}|}{\cite{N21,U20,F30}} & 3 &  \\
\cline{3-5}
&& VSM&\multicolumn{1}{p{7cm}|}{\cite{N13,N51,N53,F37}}& 4 & \\
\cline{3-5}
&& BLR &\multicolumn{1}{p{7cm}|}{\cite{N21}}& 1 & \multirow{-4}{*}{7}\\
\midrule

\multirow{17}{*}{DL-based } & \multirow{9}{*}{Sequence-based}
& DNN &\multicolumn{1}{p{7cm}|}{\cite{N13,F41,F53}}&3&\\
\cline{3-5}
&& CNN &\multicolumn{1}{p{7cm}|}{\cite{U01,N40,N61,F3,F20}}& 5&\\
\cline{3-5}
&& CNN-attention&\multicolumn{1}{p{7cm}|}{\cite{N15}}&1&\\
\cline{3-5}
&& LSTM &\multicolumn{1}{p{7cm}|}{\cite{U02,F8,F52,U04}}& 4 &\\
\cline{3-5}
&& Bi-LSTM&\multicolumn{1}{p{7cm}|}{\cite{N22,N27,N28,U11,F40}}&5&45 \\
\cline{3-5}
&& LSTM-attention&\multicolumn{1}{p{7cm}|}{\cite{N8,N22,N25,N26}}&4&\\
\cline{3-5}
&& GRU  &\multicolumn{1}{p{7cm}|}{\cite{N35,N36,N38,N64,U19,F6,F40}}& 7 & \\
\cline{3-5}
&& \multirow{2}{*}{Transformer}  &\multicolumn{1}{p{7cm}|}{\cite{N2,N7,N14,N18,N20,N23,N24,N29,N33,N37,N42,N56,N62,N66,F19,F20,F29,F40,F47,F51,F53}}& \multirow{2}{*}{21} & \\
\cline{2-6}

&\multirow{3}{*}{Tree-based}
& Tree-lstm &\multicolumn{1}{p{7cm}|}{\cite{N40,N48}} & 2 & \\ 
\cline{3-5}
&& Tree-Transformer  &\multicolumn{1}{p{7cm}|}{\cite{N58} }& 1 & 4 \\
\cline{3-5}
&& Tree-GRU  &\multicolumn{1}{p{7cm}|}{\cite{F47} }& 1 & \\
\cline{2-6}

&\multirow{5}{*}{GNN-based}
& CFGNN &\multicolumn{1}{p{7cm}|}{\cite{N49}} &1 & \\
\cline{3-5}
&& GAT &\multicolumn{1}{p{7cm}|}{\cite{U11,U19,F3,F23,F38}}&5 & \\
\cline{3-5}
&& GCN &\multicolumn{1}{p{7cm}|}{\cite{U03,N27,F20,F23,F57}}&5 &11 \\ 
\cline{3-5}
&& GGNN  & \multicolumn{1}{p{7cm}|}{\cite{N49}} & 1 &  \\
\cline{3-5}
&& DH-CNN &\multicolumn{1}{p{7cm}|}{\cite{F48}}& 1 & \\ 
\midrule
\multirow{2}{*}{LLM-based} & \multirow{2}{*}{-- }
&\multirow{2}{*}{-- } & \multicolumn{1}{p{7cm}|}{\cite{F1,F10,F11,F12,F15,F16,F17,F18,F2,F26,F27,F28,F33,F35,F39,F4,F40,F44,F46,F50,F52,F54,F59,F7,F9,N1,N3,N30,N31,N4,N54,N6}}& \multirow{2}{*}{32}  & \multirow{2}{*}{32} 
\\
\bottomrule
\end{tabular}
\end{table}

In code intelligence tasks, contextual data is modeled after preprocessing. However, traditional context-irrelevant approaches struggle to handle such contextual complexity effectively. This is due to the overwhelming scale of contextual data combined with sparse task-relevant information. Conventional methods struggle to capture the deep, hierarchical relationships and nuanced dependencies in complex repositories, failing to extract task-related information from context. These difficulties necessitate developing new methods. 

In this subsection, we focus on context-aware techniques presented in new technique papers, excluding benchmarks, case studies, and empirical studies. We identify 122 new technique papers on context-aware techniques and categorize their modeling approaches into four groups: rule, feature, deep learning, and large language models. Furthermore, if a paper implements a new context-aware method on models with more than 1 billion parameters, we classify these models as large language models rather than simply as transformer-based.

\subsubsection{Rule-Based}

Rule-driven modeling methodologies, constructed from domain-specific prior knowledge, have advantages in specialized task scenarios. Rule-based models demonstrate two advantages. First, the verifiable nature of rule systems ensures output reliability. Second, rule-based architectures maintain predictable computational complexity. 

As evidenced by 29 representative studies, this paradigm encodes expert knowledge into deterministic rules, well-suited for tasks such as program repair~\cite{N39,N55,N60,U06,U08,U09} and defect detection~\cite{N52,N57,N63,U10,U14,U16,U17}. For example, rule-based models facilitate coding intelligence tasks through predefined similarity assessment or matching protocols between existing context and target inputs. Wang et al.~\cite{N57} exemplify this approach through three rule-governed similarity dimensions: version evolution patterns, historical reports correlations, and structural code correspondences. Their defect detection system applies verifiable scoring rules to generate prioritized suspicious file rankings through weighted aggregation. Jiang et al.~\cite{N55} leverage rule-driven AST context analysis to identify semantically code patterns relevant to historical patches. Their methodology systematically mines potential defective source code from existing code repositories, distilling recurrent repair templates through structural AST context rule matching. 

Beyond program repair and defect detection, rule-based modeling is applicable to other code intelligence tasks. For instance, Wong et al.~\cite{N59} propose a rule-based model to retrieve similar comments for code summarization. This method applies semantic filtering rules to eliminate irrelevant comments that contain project-specific identifiers or non-generalizable implementation details. Zhang et al.~\cite{N12} build a rule-based system for continuous clone detection during software evolution. This system continuously tracks code modifications and analyzes contextual patterns in source code evolution. Through dynamic rule evaluation, this system is suitable for identifying clone occurrences and classifying clone types, subsequently providing real-time notifications or warnings to developers for repository refinement.

Compared with machine learning models, rule-based systems provide deterministic reasoning mechanisms that ensure computational traceability and execution efficiency. However, these systems necessitate extensive domain knowledge engineering and demonstrate limited adaptability to cross-domain scenarios, as their performance is intrinsically tied to the completeness and precision of manually crafted rules.

\subsubsection{Feature-Based}

Feature-based modeling represents a data-driven paradigm that systematically transforms information into quantifiable feature vectors through a structured pipeline of data processing techniques. Through careful pipeline design, this methodology enables the effective extraction of meaningful features from raw contextual data to solve code intelligence tasks. 

Seven studies demonstrate the paradigm's versatility through strategic deployment of classical machine learning architectures such as support vector machine (SVM)~\cite{N21,N51,N53}, decision tree~\cite{N21,U20,F30}, vector space model (VSM)~\cite{N13,N51,N53,F37}, and bayesian logistic regression (BLR)~\cite{N21}. For example, Zhou et al.~\cite{N53} leverage bug reports to extract keywords and construct a query vector. This vector is inputted to a revised VSM to rank suspicious bug files while considering the context of previously fixed, similar bug reports. Wen et al.~\cite{F37} utilize VSM to receive context chunks extracted from code changes, achieving improved bug detection with finer granularity and better performance than baselines. Bibaev et al.~\cite{F30} employ CatBoost, a decision-tree-based model, to enhance code completion. They collect relevant features from the IDE context logs, such as uncompleted code prefixes, syntactic context, syntactic feature, and session history. These features reduce the average number of user operations while providing more effective completion suggestions than classic methods. Liu et al.~\cite{N21} employ a BLR-based classifier to analyze bug types from bug reports, subsequently implementing a parameter extraction strategy and recommending repair methods based on the extracted information.

The feature-based approaches require lower data dependency and maintain interpretability through crafted features compared with methods based on deep learning. However, reliance on manual feature engineering makes it challenging to capture complex semantic information and limits its generalizability across tasks, as domain-specific features may not readily transfer to other domains.

\subsubsection{Deep Learning-Based}

Unlike traditional machine learning algorithms, deep learning models are composed of multiple layers. This capability allows deep learning models to automatically learn hierarchical feature representations from raw data, reducing reliance on human-crafted features and expanding the potential for context modeling. Deep learning models enable us to process context in code intelligence tasks. Based on the input of models, we classify deep learning models into three categories: sequence-based, tree-based, and graph-based. 

Forty-nine papers develop new methods or propose frameworks based on sequence-based deep learning models. Early works simply model context with deep neural networks (DNNs). For example, Lam et al.~\cite{N13} develop a defect detection pipeline using two DNNs. The first DNN assesses bug relevance between bug reports and source files. The second DNN integrates bug relevance, text similarity, and project information to finalize identifying suspicious source files.

In terms of convolutional neural network (CNN), Lutellier et al.~\cite{U01} combine a context-aware neural machine translation technique (NMT) with CNN to outperform recurrent neural network (RNN) in program repair. Li et al.~\cite{N40} conceive of CNN as a reranking model to predict the best performance patch. Liu et al.~\cite{N61}, Li et al.~\cite{N15}, and Qiu et al.~\cite{F3} adapt the CNN-based model as a classifier for defect detection. Liu et al.~\cite{N61} develop a multi-channel CNN that inputs both source code features and external knowledge features. Li et al.~\cite{N15} create an attention-CNN model that processes combined vectors generated from AST, CFG, and PDG. Qiu et al.~\cite{F3} leverage a CNN model with a squeeze-and-excitation network (SENet) layer to adaptively fuse the features learned from the AST, PDG, and statements.

Designed to capture long-term dependencies in sequential data, long short-term memory (LSTM) and gated recurrent unit (GRU) are two types of RNN introduced to solve code intelligence tasks~\cite{N8,N22,N35}. LSTM uses gating mechanisms to regulate the dynamic retention and forgetting of information, preventing the vanishing or exploding gradient problems found in traditional RNNs. For example, Lin et al.~\cite{N28} treat AST context as a sequence to input LSTM and predict patch correctness. Lyu et al.~\cite{F8} construct API dependency graphs as the input of LSTM for code generation. Compared with LSTM, GRU does not have a distinct memory cell but manages memory through its hidden state and improves computational efficiency by having a more straightforward gating mechanism. These advantages help the model maintain low memory usage and computational cost when utilizing context at scale. For example, some attempts~\cite{N34,N36,N38} in code summarization employ GRU to hierarchically model context. Haque et al.~\cite{N34} use one GRU per function to embed context function while utilizing a fully connected neural network to combine features from these RNNs and predict code summaries. Bansal et al.~\cite{N38} employ similar methods by using one GRU to embed each called function and utilizing convGNN to combine the embedding and summarize the code. Bansal et al.~\cite{N36} construct a project-level embedding method for code summarization by constructing pipelines from word embedding, subroutine embedding, and file embedding to project embedding. Besides basic LSTM and GRU structures, researchers also employ bi-directional structures and attention mechanisms to improve performance. For example, RepairNet~\cite{N8} constructs an Attention-LSTM structure to receive compilation information cyclically for automatically compiling failed program repair. Yuan et al.~\cite{N27} trains a Bi-LSTM model to capture CFG features before computing feature similarity for clone detection. CoRec~\cite{N22} constructs a Bi-LSTM with an attention mechanism to model retrieved context and generate a commit message. Yu et al.~\cite{U19} and Wang et al.~\cite{F38} both use a Bi-GRU as part of the encoder and another GRU as a decoder for code comment generation. Yu et al.~\cite{U19} utilize a GAT to capture graph representations between classes, and Wang et al.~\cite{F38} employ a GAN to capture the relation between function graphs and comment graphs.

Transformer~\cite{transformer} is one of the most popular models in recent years. Multi-head attention and self-attention mechanisms enable transformers to capture long-range dependencies, making them well-suited to processing long contextual information effectively. The most common way for a transformer is to add the contextual input type. For example, TransRepair~\cite{N2} considers the inputs of diagnostic feedback from the compiler and source code context extracted by the lexical analyzer to repair the broken code. Washio and Miyao~\cite{N7} extract API document information and serve it as input for a transformer-based code generator. Many studies focus on precisely capturing relevant context before inputting the transformer, such as based on static analysis~\cite{N14} or retrieval~\cite{N29,F29}. Other studies, however, focus on changing attention mechanisms. For example, LongCoder~\cite{N62} proposes window attention, global attention, and bridge attention to better capture source code context for code completion.

To better model tree-based context, researchers also propose tree-based deep learning models for code intelligence tasks 
~\cite{N40,N48,N58,F47}. Basts~\cite{N48} proposes a block-wise AST splitting method using CFG and fully utilizes rich tree-form syntax structure in AST by inputting a tree-LSTM model. DLFix~\cite{N40} proposes a two-tier DL model for automated program repair. The first layer utilizes a tree-LSTM to learn the context of bug fixes, and its output serves as an additional weighting input, which provides context information for the second layer. Gao et al.~\cite{N58} deploy a tree-transformer structure to capture local symbolic information and global syntactic structure for code summarization. Zhao et al.~\cite{F47} utilize attention tree-GRU to process diff code AST and provide rich semantic representation for fine-grained bug localization.

The GNN-based model is suitable for modeling graph-based contexts by receiving messages from other nodes; nodes aggregate and update features to effectively present the graph's structure. Zhang et al.~\cite{N49} construct a CFG-based graph neural network (CFGNN) to detect condition-related bugs, containing a graph-structured LSTM and an API-usage attention mechanism. Regarding graph attention network (GAT), Yu et al.~\cite{U11} design a parallel node internal multi-head attention layer to capture key tokens while utilizing GAT and edge contraction pooling to capture critical statements and edges. This model outperforms the state-of-the-art (SOTA) lightweight model with over 20x faster detection. To better model graph-based context, Yu et al.~\cite{U19} employ GAT to model class-level contextual graph, Wang et al.~\cite{F38} adopt GAT to model function graph and comment graph, Qiu et al.~\cite{F3} propose to use GAT to model the combination of AST and PDG. Researchers also implement graph convolutional networks (GCNs) to process context, such as CFG~\cite{N27}, flow-augmented graph~\cite{U03} (derived from CFG and DFG), and delta graph~\cite{F20} (derived from CPG). Besides GAT and GCN, Xu et al.~\cite{F23} build up a GNN classifier framework to predict defects and implement this framework in 5 different GNNs: GAT GCN, graph isomorphism network (GIN), simplified graph convolution (SGC) network, and GraphSAGE (SAGE). Haque et al.~\cite{N34} embeds AST with graph-NN, Zhang et al.~\cite{N49} utilize gated graph sequence neural networks (GGNN) to embed CFG, and Abdu et al.~\cite{F48} propose a deep hierarchical convolutional neural network (DH-CNN) to process context such as CFG and AST.

\subsubsection{Large Language Model-Based}

The emergence of LLMs changes the paradigm of context utilization in the code intelligence domain and shows a promising future in this area. Crucially, LLM-based methods represent a fundamental shift in context handling compared to earlier deep learning architectures. The primary contrast lies in the mechanism of context integration.

Prior deep learning models integrate context through architectural modifications explicitly designed to model contexts~\cite{N28,N34,N36,N38,N39}. This approach is highly effective but demands specialized engineering for each context type. In contrast, LLM-based methods leverage context augmentation via prompt Engineering and retrieval~\cite{F1,F2,F4,F7}. They convert diverse context types into a sequential augmented prompt. This shift means that the focus moves from modeling the context structure (DL-based) to retrieving and formulating the most relevant context (LLM-based). Usually, studies implement their methods on LLMs, such as StarCoder~\cite{N54,F2,F11,F26,F27,F28,F59}, Code-Llama~\cite{F4,F11,F15,F16,F26,F28,F39,F59}, CodeGen~\cite{N31,N16,F7}, DeepSeek-Coder~\cite{F11,F26,F28,F59,F50} and GPT-3.5-Turbo~\cite{N31,F1,F4,F9,F10,F12,F16,F18,F33,F39,F44,F50,F54}, and they use vanilla LLMs as a baseline to evaluate the effectiveness of the methods. These studies can be divided into three categories: retrieval-based~\cite{N6,F2,F3}, static-analysis-based~\cite{N30,F9,F10,F59}, and iterative-based~\cite{N31,N54,F4,F7}.

The basic pipeline of retrieval-based methods extracts relevant context from pre-built repositories. The context is merged with the task to form an augmented prompt that is subsequently input into a LLM to generate the answer. Among retrieval methods, for example, DocPrompting~\cite{N6} utilizes the BM25 algorithm to retrieve API documents. RepoFormer~\cite{F2} finetunes StarCoder~\cite{li2023starcoder} to determine whether the task requires retrieval. RepoHyper~\cite{F1} constructs a repo-level semantic graph for global context representation and retrieval. 

Static analysis is another type of method specialized in collecting code-based context. By scrutinizing the code statically without compiling, methods based on static analysis extract context and build up prompts. Usually, static analysis supplies more fine-grained context compared with retrieval. For instance, RLPG~\cite{N30} proposes ten context sources of prompt and constructs a classifier to determine which prompt source should be inputted to LLM based on the task. RLCE~\cite{F9} implements a static code analysis tool capable of automatically parsing the repository into code segments based on its structure and collects relevant context based on error information. LANCE~\cite{F10} proposes a lightweight and language-agnostic static analysis to extract context. 

Iterative-based methods include methods based on retrieval and static analysis, though they are implemented through an iterative pipeline. To utilize iterative-based frameworks, RepoCoder~\cite{N31} combines the generation based on retrieval to further iteratively retrieve and generate. RepoMinCoder~\cite{F7}, based on RepoCoder, adds an information-loss-based method using minimum description length to select and rerank retrieved context. CompVPD~\cite{N54} introduces iterative human validation to generate a precise context. CocoGen~\cite{F4} iteratively retrieves context based on compiler feedback. 

\begin{center}
    \begin{myboxc}{\textbf{\ding{43} RQ3 Summary: }
    \begin{enumerate} 
        \item We summarize the data preprocessing steps discussed in the reviewed papers. The most commonly used method is to split the identifiers and perform text normalization on the source code.
        \item We analyze the structure of context-aware models, finding that sequence-based deep learning models are the most widely used. Additionally, researchers prefer employing tree-based and graph-based models to process specific types of context, such as abstract syntax trees (ASTs) and control flow graphs (CFGs).
        \item There is a trend (32 studies) exploring the usage of LLMs to process context, presenting significant opportunities for further research in this area. 
    \end{enumerate}
    }
    \end{myboxc}
\end{center}
\section{RQ4: How Are Context-Aware Code Intelligence Models Evaluated?}
\label{RQ4}
To assess the evaluation and reuse of resources in context-relevant studies, we examine the metrics used to measure model performance, the benchmarks developed to validate context-aware models, and the open-source status of the papers.

\subsection{Metric}

The evaluation metric is crucial for measuring prediction accuracy and assessing the model's performance. Table~\ref{tab:Metric} summarizes the metrics utilized in all 146 papers and details their respective usage scenarios. We classify metrics based on their function to maintain the table’s clarity and focus. Besides, we exclude some simple evaluation metrics, such as memory usage~\cite{N27} and time cost~\cite{N11}. 

Ranking metrics assess models' effectiveness in recommending solutions for code intelligence tasks. The top@k metric~\cite{U17,N50,F24,F37,F47} measures the proportion of relevant results among the top k recommendations, while MAP~\cite{U14,N13,N52,N53,N57}, MAR~\cite{U17,N51}, MFR~\cite{U17}, and MRR~\cite{U12,U14,N13,F37,F47} offer a more detailed evaluation of the ranking quality.

Classification metrics measure model performance using confusion matrices, which categorize predictions into positive and negative as one dimension and the correctness of the model prediction as another. Accuracy~\cite{N6,N8,N13,N17,N21}, precision~\cite{F41,F43,F53,F54,F57,F60}, recall~\cite{F3,F11,F12,F13,F14} and F1 score~\cite{U15,N16,N17,N21,N25} are four common classification metrics derived from confusion matrices. Matthews correlation coefficient (MCC)~\cite{N54,F3,F22,F40} further considers imbalanced datasets and utilizes confusion matrices to summarize classification quality.

Similarity metrics assess how closely the model's predictions match the ground truth. Bleu~\cite{Kishore2002Bleu}, developed from natural language processing, assesses code quality using N-gram precision and a brevity penalty. However, Bleu performs poorly in evaluating code semantics. METEOR~\cite{banerjee2005meteor} addresses this shortcoming by incorporating external knowledge sources, while CodeBLEU~\cite{ren2020codebleu} improves alignment with human judgment by adding dataflow and AST analysis. Additionally, the application of UML to code summarization~\cite{N35} introduces  CIDEr~\cite{vedantam2015cider}, a metric from computer vision that combines BLEU with vector space modeling to assess whether the model captures crucial information from UML. Besides, metrics such as BM25 score~\cite{N28} and USE~\cite{N9,N37,N38} also utilize algorithms or models to judge similarity.

Model-related metrics contain perplexity, Rlmp, and area under the curve (AUC). Perplexity~\cite{U01,U04,F29} quantifies how well a probability model predicts a sample by measuring the average uncertainty or ``surprise of the model when encountering new data. Rlmp~\cite{U17,F39} evaluates the improvement of a proposed model compared with baseline models. AUC~\cite{U02,N54,F23,F40}, a widely used metric for evaluating the performance of binary classification models, measures the model's ability to distinguish between positive and negative classes by calculating the area under the receiver operating characteristic (ROC) curve. 

Besides, compiler-based metrics use compilers to assess the quality of predicted code, such as by checking whether the code compiles~\cite {N1,F51} or passes test cases~\cite{N3,N4,N5,N6,N28,F4,F5}. Frequency-based metrics measure how often certain elements or patterns occur, such as normalized inverse word frequency~\cite{F6} or unchanged identifiers ratio~\cite{N10}. Coverage-based metrics evaluate how thoroughly a model considers relevant conditions, such as inspecting necessary statements~\cite{N60}, or how well the generated outputs cover expected elements, such as including libraries~\cite{F12}, APIs~\cite{F55}, or dependencies~\cite{F15}. Higher coverage typically indicates a more comprehensive evaluation. Task-specific metrics are designed to fit specific code intelligence tasks, such as the full repair metric~\cite{N2} or the fix location filtering score~\cite{U08}.

\newpage
\begin{center}
    \scriptsize
    \renewcommand\arraystretch{1.8}
    \setlength{\LTpre}{0pt}
    \setlength{\LTpost}{0pt}
    \belowrulesep=0pt
    \aboverulesep=0pt
    \setlength{\tabcolsep}{2pt}
    \begin{longtable}{|p{1.7cm}|p{2.1cm}|p{5.2cm}|p{3cm}|p{3cm}|}
    \caption{The Metrics of Seven Types of Code Intelligence Tasks}
    \label{tab:Metric} 
    \\
    \hline
    \textbf{Types} &\textbf{Metrics} &\textbf{Meaning} &\textbf{Relevant Papers} &\textbf{Tasks}
    \\
    \hline
    \multirow{8.25}{*}{ \textbf{Ranking}}
    &\multirow{1.5}{*}{Top@K (Hit@K)}
    &This metric measures the ability of a model to include  
    relevant items in the top K recommendations.
    &\cite{U17,N50,N51,N52,N53,N58,N60,N63,N66,F24,F37,F47}
    &Defect Detection, Code Generation, Program Repair
    \\
    \cline{2-5}
    &\multirow{2.25}{*}{MAP (@k)}
    &Mean Average Precision (MAP) calculates the average precision for each query for each class. MAP@k calculates the same but within the top k results. 
    &\cite{U14,N13,N52,N53,N57,N63,F24,F29,F37,F47}
    &Defect Detection
    \\
    \cline{2-5}
    &\multirow{1.5}{*}{MAR}
    &Mean Average Rank (MAR) denotes the mean of the average rank of finding all items.
    &\cite{U17,N51}
    &Defect Detection
    \\
    \cline{2-5}
    &\multirow{1.5}{*}{MFR}
    &Mean First Rank (MFR) is the mean of localizing the first faulty statement's rank of all faults under multiple faulty scenarios. 
    &\cite{U17}
    &Defect Detection
    \\
    \cline{2-5}
    &\multirow{1.5}{*}{MRR}
    &Mean reciprocal rank (MRR) measures the average of the reciprocal ranks of the first relevant result for a set of queries.
    &\cite{U12,U14,N13,N50,N51,N52,N53,N57,N63,F37,F47}
    &Code Completion, Defect Detection
    \\
    \midrule
    \multirow{13.25}{*}{Classification}&\multirow{2.25}{*}{Accuracy}
    &Accuracy represents the proportion of correctly classified samples to the total number of samples.   &\cite{U02,U03,U04,U12,U20,N1,N2,N6,N8,N13,N17,N21,N32,N56,N61,F1,F3,F8,F10,F11,F12,F21,F23,F40,F57}
    &Program Repair, Code Completion, Code Generation, Defect Detection
    \\
    \cline{2-5}
    &\multirow{3}{*}{Precision (@k)}
    &Precision measures the proportion of correctly predicted positive instances out of all instances predicted as positive. Precision @k calculates the same, but within the top k results. 
    &\cite{U07,N15,U02,U06,U11,U12,U15,N21,N25,N27,N42,N43,N44,N49,N54,N56,N61,N66,F3,F11,F12,F22,F23,F29,F38,F41,F43,F53,F54,F57,F60}
    &Code Generation, Program Repair, Code Completion, Clone Detection, Defect Detection
    \\
    \cline{2-5}
    &\multirow{3}{*}{Recall (@k)}
    &Recall represents the proportion of correctly predicted positive samples among all positive samples.   Recall@K measures the same, but within the top k suggestions instead of positive samples.
    &\cite{U07,N15,U02,U11,U12,U14,U15,U16,N17,N21,N25,N27,N42,N43,N44,N49,N54,N56,N61,N66,F3,F11,F12,F13,F14,F22,F30,F23,F41,F43,F53,F54,F57}
    &Code Generation, Program Repair, Code Completion, Defect Detection, Clone Detection
    \\
    \cline{2-5}
    &\multirow{3}{*}{F1-score}
    &F1-score comprehensively considers the model's ability by measuring the harmonic mean of precision and recall.
    &\cite{N66,N15,N6,U02,U11,U12,U15,N16,N17,N21,N25,N27,N42,N43,N49,N54,N56,N61,N65,F8,F11,F12,F16,F22,F23,F40,F41,F43,F53,F57,F59}
    &Code Generation, Code Completion, Code Summarization, Program Repair, Clone Detection, Defect Detection
    \\
    \cline{2-5}
    &\multirow{3}{*}{MCC}
    &Matthews correlation coefficient (MCC) evaluates binary classification performance using the confusion matrix. This metric provides a balanced score even when the class sizes are imbalanced. 
    &\cite{N54,F3,F22,F40}
    &Defect Detection
    \\
    \midrule
    \multirow{24}{*}{\textbf{Similarity}}
    &\multirow{3}{*}{BLEU}
    &BLEU computes the similarity between prediction and ground truth sequences by assessing the n-gram overlap. &\cite{U07,U03,U19,N6,N7,N9,N16,N18,N20,N22,N23,N24,N26,N29,N34,N35,N36,N37,N38,N41,N48,N58,N64,F6,F8,F15,F19,F20,F31,F38,F42,F44,F45,F46,F48,F52}
    &Code Generation, Code Summarization, Program Repair, Commit Message Generation, Code Completion
    \\
    \cline{2-5}
    &\multirow{3}{*}{CodeBLEU}
    &CodeBLEU is a bleu-modified metric tailored for code-related tasks. CodeBLEU evaluates the similarity of code by weighting the combination of n-gram match, data-flow match, and AST match.
    &\cite{N66,F1,F15,F17,F18}
    &Code Generation, Code Completion
    \\
    \cline{2-5}
    &\multirow{2.25}{*}{EM}
    &Exact Match (EM) measures the proportion of predictions that match the ground truth exactly.
    &\cite{N2,N6,N14,N29,N30,N31,N33,N62,N65,N66,U07,F1,F2,F7,F10,F11,F15,F16,F17,F18,F26,F27,F28,F29,F44,F49,F59}
    &Code Generation, Code Completion, Commit Message Generation
    \\
    \cline{2-5}
    &\multirow{2.25}{*}{ES}
    &Edit Similarity (ES) measures the similarity between two strings, defined as the minimum number of single-character edits.
    &\cite{N14,N30,N31,N62,N65,F2,F7,F11,F15,F16,F17,F26,F27,F28,F29,F44,F51,F59}
    &Code Generation, Code Completion, Commit Message Generation
    \\
    \cline{2-5}
    &\multirow{1.5}{*}{SM}
    &Syntax Match (SM) quantifies the match of subtrees within the code.
    &~\cite{F18}
    &Code Completion
    \\
    \cline{2-5}
    &\multirow{2.25}{*}{ROUGE}
    &ROUGE utilizes the longest common subsequence to measure the matching degree between the generated and reference results.    &\cite{U19,N18,N20,N22,N23,N24,N25,N26,N34,N35,N36,N38,N41,N48,N58,N64,F6,F8,F19,F20,F38,F42,F45,F48,F52}
    &Code Generation, Code Summarization, Commit Message Generation, Program Repair
    \\
    \cline{2-5}
    &\multirow{2.25}{*}{METEOR}
    &METEOR computes the harmonic mean between unigram precision and recall.    &\cite{N9,N16,N18,N20,N22,N23,N24,N25,N26,N35,N37,N38,N41,N58,N64,F6,F19,F20,F31,F42,F46,F48,F52,F57}
    &Code Summarization, Commit Message Generation, Program Repair
    \\
    \cline{2-5}
    &\multirow{1.5}{*}{USE}
    &The metric encodes summaries into vectors and computes the distance between predicted and reference summary vectors.
    &\cite{N9,N37,N38}
    &Code Summarization
    \\
    \cline{2-5}
    &\multirow{2.25}{*}{CIDEr}
    &CIDEr is a consensus-based evaluation metric used in image captioning tasks to measure whether the model captures important messages.
    &\cite{N35,F8}
    &Code Generation, Code Summarization
    \\
    \cline{2-5}
    &\multirow{2.25}{*}{BM25 score}
    &Best Matching 25 (BM25) score estimates the relevance of texts based on term frequency, inverse document frequency, and document length normalization. 
    &\cite{N28}
    &Code Generation
    \\
    \cline{2-5}
    &\multirow{1.5}{*}{RIBES}
    &RIBES estimates models based on rank correlation and evaluates the distance between predictions and references.
    &\cite{F8}
    &Code Generation
    \\
    \hline
    \multirow{6}{*}{\textbf{Model-Related}}
    &\multirow{2.25}{*}{Perplexity}
    & Perplexity measures how well a probability distribution predicts a sample. This metric computes the exponentiation of the average negative log-likelihood of the predicted probabilities.
    &\cite{U01,U04,F29}
    &Program Repair, Code Completion
    \\
    \cline{2-5}
    
    &\multirow{1.5}{*}{RImp} 
    &Relative Improvement (RImp) calculates the improvement of solving tasks between the old and new techniques.
    &\cite{U17,F39}
    &Defect Detection, Program Repair
    \\
    \cline{2-5}
    &\multirow{2.25}{*}{AUC}
    &Area Under Curve (AUC) measures the effectiveness of classification models by calculating the area under the 
    receiver operating characteristic curve. 
    &\cite{U02,N54,F23,F40}
    &Program Repair, Defect Detection
    \\
    \hline
    \multirow{6}{*}{\textbf{Compiler-Based}}
    &\multirow{1.5}{*}{Pass@k}
    &This metric calculates the proportion of solved questions that pass the unit test.
    &\cite{N3,N4,N5,N6,N28,F4,F5,F13,F14,F15,F33,F34,F36,F50,F56}
    &Code Generation, Code Completion
    \\
    \cline{2-5}
    &\multirow{1.5}{*}{CR}
    &Compilation Rate (CR) measures whether the complete repository compiles successfully.
    &\cite{N1,F51}
    &Code Completion, Code Generation,
    \\
    \cline{2-5}
    &\multirow{3}{*}{ValRate}
    &Static Validity Rate (ValRate) calculates the performance of models in reducing dependency errors by evaluating the proportion of generated code that successfully passes a static check for dependency errors.
    &\cite{F15}
    &Code Generation
    \\
    \hline
    \multirow{3.75}{*}{\textbf{Frequency}}
    &\multirow{2.25}{*}{NIWF}
    &Normalized inverse word frequency (NIWF) calculates the maximum inverse word frequency (IWF) of all the tokens in a given comment.
    &\cite{F6}
    &Code Summarization
    \\
    \cline{2-5}
    &\multirow{1.5}{*}{UnchangedRatio}
    &UnchangedRatio measures whether programmers consistently change identifiers when copying and pasting code.
    &\cite{N10}
    &Program Repair
    \\
    \hline
    \multirow{8.25}{*}{\textbf{Coverage}}
    &\multirow{1.5}{*}{EXAM}
    &Exam measures the fraction of ranked statements that must be inspected before encountering a buggy statement.
    &\cite{N60}
    &Program Repair
    \\
    \cline{2-5}
    &\multirow{1.5}{*}{ACR}
    &Average Coverage Ratio (ACR) calculates the average ratio of successfully predicted statements in all buggy statements. 
    &\cite{F24}
    &Defect Detection
    \\
    \cline{2-5}
    &\multirow{1.5}{*}{Library Coverage}
    &This metric calculates the ratio of libraries already preinstalled in the repository and used in the generated code.
    &\cite{F12}
    &Code Completion 
    \\
    \cline{2-5}
    &\multirow{2.25}{*}{Dependency Coverage}
    &Dependency Coverage means the proportion of dependencies that appear in ground-truth functions while covered by the prediction.
    &\cite{F15}
    &Code Generation
    \\
    \cline{2-5}
    &\multirow{1.5}{*}{API Coverage}
    &API Coverage reports the ratio of API coverage in code snippets.
    &\cite{F55}
    &Defect Detection
    \\
    \hline
    \multirow{9}{*}{\textbf{Task-Specific}}
    &\multirow{2.25}{*}{Full Repair}
    &Full Repair evaluates the ability to fix a broken program, including the location of an error statement and the fix of this statement. 
    &\cite{N2}
    &Program Repair
    \\
    \cline{2-5}
    &\multirow{1.5}{*}{MCR}
    &Matched Change Ratio (MCR) represents the portion of changes found by a change selection method.
    &\cite{U08}
    &Program Repair
    \\
    \cline{2-5}
    &\multirow{1.5}{*}{FLFscore}
    &Fix Location Filtering Score (FLFscore) measures the model's accuracy in filtering out code that does not require fixing.
    &\cite{U08}
    &Program Repair
    \\
    \cline{2-5}
    &\multirow{1.5}{*}{DIR}
    &Dependency Invocation Rate (DIR) calculates the ratio of code dependencies successfully integrated into prediction.
    &\cite{F32}
    &Code Generation
    \\
    \cline{2-5}
    &\multirow{2.25}{*}{$p_{opt}$}
    &$p_{opt}$ measures the performance of models by comparing them with an optimal model, which orders all the modules to decrease the defect score. 
    &\cite{F25}
    &Defect Detection
    \\
    \hline
    \end{longtable}
\end{center}

However, due to the lack of suitable metrics or the need for human evaluation, studies~\cite{U05,U18,N12,N45,N46,N59,F9} use human evaluation to assess the effectiveness of their methods. For instance, to evaluate the effectiveness of notifications for continuous clone monitoring, study~\cite{N12} collects feedback from 9 participants to determine if the notifications meet their requirements. Li et al.~\cite{F42} enroll participants to evaluate the quality of commit messages from four dimensions: rationality, comprehensiveness, conciseness, and expressiveness. Developing automated metrics that better align with relevant studies may enhance the evaluation process and provide more consistent, scalable assessments.

Furthermore, current evaluation metrics predominantly focus on end-to-end performance while paying insufficient attention to context utilization. Neglecting to assess the intermediate steps of context utilization renders these methods a ``black box". This oversight can obscure potential bottlenecks or errors, ultimately complicating the optimization process. To address this, developing more fine-grained metrics for context utilization is necessary. For example, RepoExec~\cite{F32} proposes ``Dependency Invocation Rate'' to directly measure how well an LLM utilizes the provided source code context to generate code with dependencies.

\begin{sidewaystable}
    \vspace*{15cm}
    \renewcommand{\arraystretch}{2.5}
    \setlength{\tabcolsep}{0.8pt}  
    \centering
    \tiny
    \caption{Datasets created by Context-Relevant Papers. 
    ``\textcolor{green}{\ding{51}}'' indicates that datasets provide this type of context directly. ``\textcolor{blue}{$\circ$}'' means that datasets do not store this type of context but can provide it through a series of processes. ``\textcolor{red}{-}'' means that this type of context is difficult or nearly impossible to obtain from these datasets.}
    \label{tab:dataset}
    \begin{tabular}{|c|c|c|c|c|c|c|c|c|c|c|c|c|c|c|c|}
        \hline
        \multirow{4}{*}{\textbf{\makecell[c]{Tasks}}} & \multirow{4}{*}{\textbf{\makecell[c]{Languages}}} & \multicolumn{6}{c|}{\textbf{Direct Context}} & \multicolumn{8}{c|}{\textbf{Indirect Context}} \\ \cline{3-16}
        &  & \multicolumn{2}{c|}{\textbf{Code-Based}} & 
        \multicolumn{4}{c|}{\textbf{Text-Based}} &
        \multicolumn{1}{c|}{\textbf{Tree-Based}} & \multicolumn{5}{c|}{\textbf{Graph-Based}} & \multicolumn{2}{c|}{\textbf{Sequence-Based}} \\ \cline{3-16}
        &  
        & \textbf{\makecell[c]{Source\\Code}} 
        & \textbf{\makecell[c]{Code\\Diff}} & \textbf{\makecell[c]{API\\Document}} & \textbf{\makecell[c]{Bug\\Report}} & \textbf{\makecell[c]{Code\\Comment}} & \textbf{\makecell[c]{Commit\\ Message}} & \textbf{AST} & \textbf{CFG} & \textbf{DFG} & \textbf{CPG} & \textbf{PDG} & \textbf{UML} & \textbf{\makecell[c]{Compilation\\Info}} & \textbf{IDE} \\ \hline
        
        \multirow{3}{*}{\textbf{\makecell[c]{Code\\Generation}}} 
        & Python 
        &\makecell[l]{\textcolor{green}{\ding{51}}~\cite{N5,U07,F5}\\~\cite{F12,F13,F14,F32}\\~\cite{F33,F34,F56}}
        &\makecell[l]{\textcolor{green}{\ding{51}}~\cite{F34}}
        &\makecell[l]{\textcolor{green}{\ding{51}}~\cite{N3,N4}\\~\cite{F54}}
        & \textcolor{red}{-}
        &\makecell[l]{\textcolor{green}{\ding{51}}~\cite{N5,F5,F12}\\~\cite{F14,F32,F34,F56}}
        &\makecell[l]{\textcolor{blue}{$\circ$}~\cite{F34}}
        &\makecell[l]{\textcolor{blue}{$\circ$}~\cite{N5,U07,F5}\\~\cite{F12,F13,F14}\\~\cite{F32,F34,F56}}
        &\makecell[l]{\textcolor{blue}{$\circ$}~\cite{N5,U07,F5}\\~\cite{F12,F13,F14}\\~\cite{F32,F34,F56}}
        & \makecell[l]{\textcolor{blue}{$\circ$}~\cite{N5,U07,F5}\\~\cite{F12,F13,F14}\\~\cite{F32,F34,F56}}
        & \makecell[l]{\textcolor{blue}{$\circ$}~\cite{N5,F5,F12}\\~\cite{F14,F32,F34,F56}}
        & \makecell[l]{\textcolor{blue}{$\circ$}~\cite{N5,F5,F12}\\~\cite{F14,F32,F34,F56}}
        & \textcolor{red}{-}
        & \makecell[l]{\textcolor{blue}{$\circ$}~\cite{N3,N4,N5}\\~\cite{F5,F13,F14,F32}\\~\cite{F33,F34,F56}}
        & \textcolor{red}{-}
        \\ 
        \cline{2-16} 
        & Java   
        & \makecell[l]{\textcolor{green}{\ding{51}}~\cite{N5,N30,F5}}
        & \textcolor{red}{-}
        & \textcolor{red}{-}
        & \textcolor{red}{-}
        & \makecell[l]{\textcolor{green}{\ding{51}}~\cite{N5,F5}}
        & \textcolor{red}{-}
        & \makecell[l]{\textcolor{blue}{$\circ$}~\cite{N5,N30,F5}}
        & \makecell[l]{\textcolor{blue}{$\circ$}~\cite{N5,N30,F5}}
        & \makecell[l]{\textcolor{blue}{$\circ$}~\cite{N5,N30,F5}}
        & \makecell[l]{\textcolor{blue}{$\circ$}~\cite{N5,N30,F5}}
        & \makecell[l]{\textcolor{blue}{$\circ$}~\cite{N5,N30,F5}}
        & \textcolor{red}{-}
        & \makecell[l]{\textcolor{blue}{$\circ$}~\cite{N5,F5}}
        & \textcolor{red}{-}
        \\ \cline{2-16} 
        & Other  
        & \textcolor{red}{-}
        & \textcolor{red}{-}
        &\makecell[l]{\textcolor{green}{\ding{51}}~\cite{N6}}
        & \textcolor{red}{-}
        & \textcolor{red}{-}
        & \textcolor{red}{-}
        & \textcolor{red}{-}
        & \textcolor{red}{-}
        & \textcolor{red}{-}
        & \textcolor{red}{-}
        & \textcolor{red}{-}
        & \textcolor{red}{-}
        & \textcolor{red}{-}
        & \textcolor{red}{-}
        \\ \hline
        
        \multirow{3}{*}{\textbf{\makecell[c]{Code\\Completion}}} 
        & Python 
        & \makecell[l]{\textcolor{green}{\ding{51}}~\cite{N14,N29,N31}\\~\cite{N32,N33,N65,F10}\\~\cite{F16,F17,F27,F28}\\~\cite{F36}}
        & \textcolor{red}{-}
        &\textcolor{red}{-}
        & \textcolor{red}{-}
        & \makecell[l]{\textcolor{green}{\ding{51}}~\cite{N14,N29,N31}\\~\cite{N32,N65,F10,F16}\\~\cite{F17,F27,F28,F36}}
        & \textcolor{red}{-}
        & \makecell[l]{\textcolor{green}{\ding{51}}~\cite{U20}\\\textcolor{blue}{$\circ$}~\cite{N14,N29,N31}\\~\cite{N32,N33,N65,F10}\\~\cite{F16,F17,F27,F28}\\~\cite{F36}}
        &  \makecell[l]{\textcolor{blue}{$\circ$}~\cite{N14,N29,N31}\\~\cite{N32,N33,N65,F10}\\~\cite{F16,F17,F27,F28}\\~\cite{F36}}
        &  \makecell[l]{\textcolor{blue}{$\circ$}~\cite{N14,N29,N31}\\~\cite{N32,N33,N65,F10}\\~\cite{F16,F17,F27,F28}\\~\cite{F36}}
        & \makecell[l]{\textcolor{blue}{$\circ$}~\cite{N14,N29,N31}\\~\cite{N32,N33,N65,F10}\\~\cite{F17,F27,F28,F36}}
        & \makecell[l]{\textcolor{blue}{$\circ$}~\cite{N14,N29,N31}\\~\cite{N32,N33,N65,F10}\\~\cite{F17,F27,F28,F36}}
        & \textcolor{red}{-}
        & \makecell[l]{\textcolor{blue}{$\circ$}~\cite{F36}}
        & \makecell[l]{\textcolor{green}{\ding{51}}~\cite{F30}}
        \\ \cline{2-16} 
        & Java   
        & \makecell[l]{\textcolor{green}{\ding{51}}~\cite{N32,N65,F10}\\~\cite{F27,F28}}
        & \textcolor{red}{-}
        & \textcolor{red}{-}
        & \textcolor{red}{-}
        & \makecell[l]{\textcolor{green}{\ding{51}}~\cite{N32,N65,F10}\\~\cite{F27,F28}}
        & \textcolor{red}{-}
        & \makecell[l]{\textcolor{blue}{$\circ$}~\cite{N32,N65,F10}\\~\cite{F27,F28}}
        & \makecell[l]{\textcolor{blue}{$\circ$}~\cite{N32,N65,F10}\\~\cite{F27,F28}}
        & \makecell[l]{\textcolor{blue}{$\circ$}~\cite{N32,N65,F10}\\~\cite{F27,F28}}
        & \makecell[l]{\textcolor{blue}{$\circ$}~\cite{N32,N65,F10}\\~\cite{F27,F28}}
        & \makecell[l]{\textcolor{blue}{$\circ$}~\cite{N32,N65,F10}\\~\cite{F27,F28}}
        & \textcolor{red}{-}
        & \textcolor{red}{-}
        & \textcolor{red}{-}
        \\ \cline{2-16} 
        & Other  
        & \makecell[l]{\textcolor{green}{\ding{51}}~\cite{N65,F28,F36}\\~\cite{F50}}
        & \textcolor{red}{-}
        & \textcolor{red}{-}
        & \textcolor{red}{-}
        & \makecell[l]{\textcolor{green}{\ding{51}}~\cite{N65,F28,F36}\\~\cite{F50}}
        & \textcolor{red}{-}
        & \makecell[l]{\textcolor{blue}{$\circ$}~\cite{N65,F28,F36}\\~\cite{F50}}
        & \makecell[l]{\textcolor{blue}{$\circ$}~\cite{N65,F28,F36}\\~\cite{F50}}
        & \makecell[l]{\textcolor{blue}{$\circ$}~\cite{N65,F28,F36}\\~\cite{F50}}
        & \makecell[l]{\textcolor{blue}{$\circ$}~\cite{N65,F28,F36}\\~\cite{F50}}
        & \makecell[l]{\textcolor{blue}{$\circ$}~\cite{N65,F28,F36}\\~\cite{F50}}
        & \textcolor{red}{-}
        & \makecell[l]{\textcolor{blue}{$\circ$}~\cite{F36,F50}}
        & \textcolor{red}{-}
        \\ 
        \hline
        
        \multirow{3}{*}{\textbf{\makecell[c]{Code\\Summarization}}} 
        & Python
        &\textcolor{red}{-}
        &\textcolor{red}{-}
        &\textcolor{red}{-}
        &\textcolor{red}{-}
        &\textcolor{red}{-}
        &\textcolor{red}{-}
        &\textcolor{red}{-} 
        &\textcolor{red}{-}
        &\textcolor{red}{-}
        &\textcolor{red}{-}
        &\textcolor{red}{-}
        &\textcolor{red}{-}
        &\textcolor{red}{-}
        &\textcolor{red}{-}
        \\ \cline{2-16} 
        & Java   
        & \makecell[l]{\textcolor{green}{\ding{51}}~\cite{N25,N26,N35}\\~\cite{F38,F52,F58}}
        & \textcolor{red}{-}
        & \makecell[l]{\textcolor{green}{\ding{51}}~\cite{N16}}
        & \textcolor{red}{-}
        & \makecell[l]{\textcolor{green}{\ding{51}}~\cite{N25,N26,N35}\\~\cite{F38,F52,F58}}
        & \textcolor{red}{-}
        & \makecell[l]{\textcolor{blue}{$\circ$}~\cite{N25,N26,N35}\\~\cite{F38,F52,F58}}
        & \makecell[l]{\textcolor{blue}{$\circ$}~\cite{N25,N26,N35}\\~\cite{F38,F52,F58}}
        & \makecell[l]{\textcolor{blue}{$\circ$}~\cite{N25,N26,N35}\\~\cite{F38,F52,F58}}
        & \makecell[l]{\textcolor{blue}{$\circ$}~\cite{F38,F52,F58}}
        & \makecell[l]{\textcolor{blue}{$\circ$}~\cite{F38,F52,F58}}
        & \makecell[l]{\textcolor{green}{\ding{51}}~\cite{N35}}
        & \textcolor{red}{-}
        & \textcolor{red}{-}
        \\ \cline{2-16} 
        & Other  
        &  \makecell[l]{\textcolor{green}{\ding{51}}~\cite{N64}}
        & \textcolor{red}{-}
        & \textcolor{red}{-}
        & \textcolor{red}{-}
        & \makecell[l]{\textcolor{green}{\ding{51}}~\cite{N64}}
        & \textcolor{red}{-}
        & \makecell[l]{\textcolor{blue}{$\circ$}~\cite{N64}}
        & \makecell[l]{\textcolor{blue}{$\circ$}~\cite{N64}}
        & \makecell[l]{\textcolor{blue}{$\circ$}~\cite{N64}}
        & \makecell[l]{\textcolor{blue}{$\circ$}~\cite{N64}}
        & \makecell[l]{\textcolor{blue}{$\circ$}~\cite{N64}}
        & \textcolor{red}{-}
        & \textcolor{red}{-}
        & \textcolor{red}{-}
        \\ \hline
        \multirow{3}{*}{\textbf{\makecell[c]{Commit\\ Message\\Generation}}}
        & Python 
        & \makecell[l]{\textcolor{green}{\ding{51}}~\cite{F19,F44}}
        & \makecell[l]{\textcolor{green}{\ding{51}}~\cite{F19,F44}}
        & \textcolor{red}{-}
        & \textcolor{red}{-}
        & \textcolor{red}{-}
        & \makecell[l]{\textcolor{green}{\ding{51}}~\cite{F19,F44}}
        & \makecell[l]{\textcolor{blue}{$\circ$}~\cite{F19,F44}}
        & \makecell[l]{\textcolor{blue}{$\circ$}~\cite{F19,F44}}
        & \makecell[l]{\textcolor{blue}{$\circ$}~\cite{F19,F44}}
        & \makecell[l]{\textcolor{blue}{$\circ$}~\cite{F19,F44}}
        & \makecell[l]{\textcolor{blue}{$\circ$}~\cite{F19,F44}}
        & \textcolor{red}{-}
        & \textcolor{red}{-}
        & \textcolor{red}{-}
        \\ \cline{2-16} 
        & Java   
        & \makecell[l]{\textcolor{green}{\ding{51}}~\cite{N18,N22,N23}\\~\cite{F44}}
        & \makecell[l]{\textcolor{green}{\ding{51}}~\cite{N18,N22}\\~\cite{N23,F44}}
        & \textcolor{red}{-}
        & \textcolor{red}{-}
        & \textcolor{red}{-}
        &  \makecell[l]{\textcolor{green}{\ding{51}}~\cite{N18,N22}\\~\cite{N23,F44}}
        &  \makecell[l]{\textcolor{blue}{$\circ$}~\cite{N18,N22,N23}\\~\cite{F44}}
        &  \makecell[l]{\textcolor{blue}{$\circ$}~\cite{N18,N22,N23}\\~\cite{F44}}
        &  \makecell[l]{\textcolor{blue}{$\circ$}~\cite{N18,N22,N23}\\~\cite{F44}}
        &  \makecell[l]{\textcolor{blue}{$\circ$}~\cite{N23,F44}}
        &  \makecell[l]{\textcolor{blue}{$\circ$}~\cite{N23,F44}}
        & \textcolor{red}{-}
        &\textcolor{red}{-}
        & \textcolor{red}{-}
        \\ \cline{2-16} 
        & Other 
        & \makecell[l]{\textcolor{green}{\ding{51}}~\cite{F44}}
        &\makecell[l]{\textcolor{green}{\ding{51}}~\cite{F44}}
        & \textcolor{red}{-}
        &\textcolor{red}{-}
        &\textcolor{red}{-}
        &\makecell[l]{\textcolor{green}{\ding{51}}~\cite{F44}}
        &\makecell[l]{\textcolor{blue}{$\circ$}~\cite{F44}}
        &\makecell[l]{\textcolor{blue}{$\circ$}~\cite{F44}}
        &\makecell[l]{\textcolor{blue}{$\circ$}~\cite{F44}}
        &\makecell[l]{\textcolor{blue}{$\circ$}~\cite{F44}}
        &\makecell[l]{\textcolor{blue}{$\circ$}~\cite{F44}}
        &\textcolor{red}{-}
        &\textcolor{red}{-}
        &\textcolor{red}{-}
        \\ \hline
        
        \multirow{3}{*}{\textbf{\makecell[c]{Clone\\ Detection}}} 
        & Python 
        &\textcolor{red}{-} 
        &\textcolor{red}{-} 
        &\textcolor{red}{-}
        \makecell[l]{\textcolor{green}{\ding{51}}~\cite{F41,F43}}
        &\textcolor{red}{-}
        &\textcolor{red}{-}
        &\textcolor{red}{-}
        &\textcolor{red}{-}
        &\textcolor{red}{-}
        &\textcolor{red}{-}
        &\textcolor{red}{-} 
        &\textcolor{red}{-}
        &\textcolor{red}{-}
        &\textcolor{red}{-}
        &\textcolor{red}{-} \\ \cline{2-16} 
        & Java   
        &\makecell[l]{\textcolor{green}{\ding{51}}~\cite{N27,N43}}
        &\textcolor{red}{-}
        &\makecell[l]{\textcolor{green}{\ding{51}}~\cite{F41,F43}}
        & \textcolor{red}{-}
        &\makecell[l]{\textcolor{green}{\ding{51}}~\cite{N27,N43}}
        & \textcolor{red}{-}
        &\makecell[l]{\textcolor{blue}{$\circ$}~\cite{N27,N43}}
        &\makecell[l]{\textcolor{blue}{$\circ$}~\cite{N27,N43}}
        &\makecell[l]{\textcolor{blue}{$\circ$}~\cite{N27,N43}}
        &\makecell[l]{\textcolor{blue}{$\circ$}~\cite{N27,N43}}
        &\makecell[l]{\textcolor{blue}{$\circ$}~\cite{N27,N43}}
        &\textcolor{red}{-}
        &\textcolor{red}{-}
        & \\ \cline{2-16} 
        & Other  
        &\textcolor{red}{-}
        &\textcolor{red}{-}
        &\makecell[l]{\textcolor{green}{\ding{51}}~\cite{F41,F43}}
        &\textcolor{red}{-}
        &\textcolor{red}{-}
        &\textcolor{red}{-}
        &\textcolor{red}{-}
        &\textcolor{red}{-}
        &\textcolor{red}{-} 
        &\textcolor{red}{-}
        &\textcolor{red}{-}
        &\textcolor{red}{-}
        &\textcolor{red}{-}
        &\textcolor{red}{-} \\ \hline
        
        \multirow{3}{*}{\textbf{\makecell[c]{Defect\\ Detection}}} 
        & Python 
        &\makecell[l]{\textcolor{green}{\ding{51}}~\cite{F21}}
        &\textcolor{red}{-}
        &\textcolor{red}{-}
        &\textcolor{red}{-}
        &\makecell[l]{\textcolor{green}{\ding{51}}~\cite{F21}}
        &\textcolor{red}{-}
        &\makecell[l]{\textcolor{blue}{$\circ$}~\cite{F21}}
        &\makecell[l]{\textcolor{blue}{$\circ$}~\cite{F21}}
        &\makecell[l]{\textcolor{blue}{$\circ$}~\cite{F21}}
        &\makecell[l]{\textcolor{blue}{$\circ$}~\cite{F21}}
        &\makecell[l]{\textcolor{blue}{$\circ$}~\cite{F21}}
        &\textcolor{red}{-}
        & \textcolor{red}{-}
        &\textcolor{red}{-}
        \\ \cline{2-16} 
        & Java   
        & \makecell[l]{\textcolor{green}{\ding{51}}~\cite{N15,N49,N53}\\~\cite{F21,F23}}
        &\textcolor{red}{-}
        &\textcolor{red}{-}
        & \makecell[l]{\textcolor{green}{\ding{51}}~\cite{N53}}
        & \makecell[l]{\textcolor{green}{\ding{51}}~\cite{N15,N49,N53}\\~\cite{F21,F23}}
        &\textcolor{red}{-}
        &\makecell[l]{\textcolor{blue}{$\circ$}~\cite{N15,N49,N53}\\~\cite{F21,F23}}
        &\makecell[l]{\textcolor{blue}{$\circ$}~\cite{N15,N49,N53}\\~\cite{F21,F23}}
        &\makecell[l]{\textcolor{blue}{$\circ$}~\cite{N15,N49,N53}\\~\cite{F21,F23}}
        &\makecell[l]{\textcolor{blue}{$\circ$}~\cite{N15,N49,N53}\\~\cite{F21,F23}}
        &\makecell[l]{\textcolor{blue}{$\circ$}~\cite{N15,N49,N53}\\~\cite{F21,F23}}
        &\textcolor{red}{-}
        &\textcolor{red}{-}
        &\textcolor{red}{-}
        \\ \cline{2-16} 
        & Other  
        & \makecell[l]{\textcolor{green}{\ding{51}}~\cite{F21,F22}}
        &\textcolor{red}{-}
        &\textcolor{red}{-}
        &\textcolor{red}{-}
        & \makecell[l]{\textcolor{green}{\ding{51}}~\cite{F21,F22}}
        &\textcolor{red}{-}
        & \makecell[l]{\textcolor{blue}{$\circ$}~\cite{F21,F22}}
        &\makecell[l]{\textcolor{blue}{$\circ$}~\cite{F21,F22}}
        &\makecell[l]{\textcolor{blue}{$\circ$}~\cite{F21,F22}}
        & \makecell[l]{\textcolor{blue}{$\circ$}~\cite{F21,F22}}
        &\makecell[l]{\textcolor{blue}{$\circ$}~\cite{F21,F22}}
        &\textcolor{red}{-}
        &\textcolor{red}{-}
        &\textcolor{red}{-}
        \\ \hline
        \multirow{3}{*}{\textbf{\makecell[c]{Program\\ Repair}}} 
        & Python 
        & \makecell[l]{\textcolor{green}{\ding{51}}~\cite{F9,F60}}
        & \makecell[l]{\textcolor{green}{\ding{51}}~\cite{F60}}
        & \textcolor{red}{-}
        & \makecell[l]{\textcolor{green}{\ding{51}}~\cite{F60}}
        & \makecell[l]{\textcolor{green}{\ding{51}}~\cite{F9,F60}}
        & \makecell[l]{\textcolor{green}{\ding{51}}~\cite{F60}}
        & \makecell[l]{\textcolor{blue}{$\circ$}~\cite{F9,F60}}
        & \makecell[l]{\textcolor{blue}{$\circ$}~\cite{F9,F60}}
        & \makecell[l]{\textcolor{blue}{$\circ$}~\cite{F9,F60}}
        & \makecell[l]{\textcolor{blue}{$\circ$}~\cite{F9,F60}}
        & \makecell[l]{\textcolor{blue}{$\circ$}~\cite{F9,F60}}
        &\textcolor{red}{-}
        & \makecell[l]{\textcolor{blue}{$\circ$}~\cite{F60}}
        &\textcolor{red}{-}
        \\ \cline{2-16} 
        & Java   
        &\textcolor{red}{-}
        & \textcolor{red}{-}
        &\textcolor{red}{-}
        &\textcolor{red}{-} 
        &\textcolor{red}{-}
        &\textcolor{red}{-} 
        &\textcolor{red}{-}
        &\textcolor{red}{-}
        &\textcolor{red}{-}
        &\textcolor{red}{-} 
        &\textcolor{red}{-}
        &\textcolor{red}{-} 
        &\textcolor{red}{-} 
        &\textcolor{red}{-}
        \\ \cline{2-16} 
        & Other  
        &\textcolor{red}{-} 
        &\textcolor{red}{-}
        &\textcolor{red}{-}
        &\textcolor{red}{-}
        &\textcolor{red}{-}
        &\textcolor{red}{-}
        &\textcolor{red}{-}
        &\textcolor{red}{-}
        &\textcolor{red}{-}
        &\textcolor{red}{-}
        &\textcolor{red}{-}
        &\textcolor{red}{-}
        &\textcolor{red}{-}
        &\textcolor{red}{-}
        \\ \hline
    \end{tabular}
\end{sidewaystable}

\begin{table}[t]
\centering
\setlength{\LTpre}{0pt}
\setlength{\LTpost}{0pt}
\footnotesize
\belowrulesep=0pt
\aboverulesep=0pt
\setlength{\tabcolsep}{1mm}
 \renewcommand{\arraystretch}{1.5}  
\caption{The Open Source Status of Seven Code Intelligence Tasks}
\begin{tabular}{|l|p{4cm}|p{3.7cm}|p{4cm}|}
\toprule
\textbf{Tasks}  
& \textbf{Data Available}  
& \textbf{Model Weight Available}   
& \textbf{Code Available} \\
\midrule
Code Generation
&\cite{N4,N5,N6,N66,U03,U07,F12,F13,F14,F32,F34,F56}
&\cite{N6,N66,U03,F8}
&\cite{N4,N5,N6,N66,U03,U07,F4,F5,F8,F12,F13,F14,F32,F34,F56}
\\ 
\hline
Code Completion
&\cite{N1,N14,N30,N31,N32,N33,N62,F16,F27,F29}
&\cite{N1,N30,N32,N33,N62,F1}
&\cite{U12,N1,N14,N30,N31,N32,N33,N62,F1,F2,F16,F26,F27,F29}
\\
\hline
Code Summarization
&\cite{N26,N34,N35,N36,N38,N42,N48,N58,N64,F31,F46,F52}
&\cite{N29,N36}
&\cite{N26,N29,N34,N36,N38,N42,N48,N58,F31,F46}
\\   
\hline
Commit Message Generation
&\cite{N18,N22,N23,F19,F21,F44}
&\cite{N23}
&\cite{N18,N20,N22,N23,F19,F21}
\\
\hline
Clone Detection 
&\cite{N43,U11,F41,F43}
& --
&\cite{N43,U11,F41,F43}
\\
\hline
Defect Detection
&\cite{N15,N49,N51,N53,N63,F23,F60}
&\cite{N15}
&\cite{U17,N13,N15,N49,N51,N53,F57,N63}
\\
\hline
Program Repair 
&\cite{U01,U02,U03,U04,N40,N60}
&\cite{U01}
&\cite{U01,U02,U03,U04,U09,N40,N55,N60}
\\ 
\bottomrule
\end{tabular}

\label{tab:openStatus}
\end{table}

\subsection{Dataset}

To support further research, we organize the datasets created in context-relevant papers in this subsection.  Table~\ref{tab:dataset} shows the specifics of datasets from three dimensions: code intelligence tasks, relevant contexts, and programming languages. We focus exclusively on datasets created within papers, excluding benchmarks or datasets used from other sources.

We find an uneven provision of context across different code intelligence tasks. This phenomenon is relevant to context utilization research in the corresponding task. For example, API documents are used in tasks such as code generation~\cite{N3,N4,F54} and clone detection~\cite{F41,F43}, creating datasets that include API documents. However, many types of context remain underutilized in code intelligence tasks, particularly for conditions that datasets do not include.

Regarding programming language selection, researchers tend to create datasets in Java or Python, with other languages less common.  Additionally, multilingual datasets remain relatively scarce, underscoring the need for more diverse datasets to enable more accurate and comprehensive assessments. We also find that programming language selection is relevant to code intelligence tasks. For example, code generation and program repair tend to build up datasets with the context in Python language~\cite{N5,U07,F5,F9,F33,F34,F56,F60}, while the code summarization task tends to create datasets with the context in Java language~\cite{N25,N26,N35,F38,F52,F58}.

As for obtaining context, it's relevant whether the context is direct or indirect but ultimately determined by dataset format. For example, datasets usually store source code context~\cite{U07,F5,F12,F33,F56} but less likely to store AST~\cite{U20}. UML~\cite{N35} and IDE usage history~\cite{F30} are indirect types of context but are obtainable as they are stored in datasets.

\subsection{Open Source Status}

Open source refers to the free availability of research data, code, model weights, and other relevant materials on public platforms. This approach promotes transparency, reproducibility, and collaboration within the research community. As shown in Table ~\ref{tab:openStatus}, we analyze the open-source status of papers in three dimensions: data, model weights, and code. ``Data'' refers to datasets or benchmarks created and used in the research, ``model weights'' denote the parameters of models proposed and trained in relevant papers, and ``code'' encompasses the implementation details of the models or tools described. Open source is essential for advancing research, as it enables the reproduction of results and fosters improvements through collaborative efforts.

Among the 146 selected papers, 71 provide open access to at least one type of resource: data, model weights, or code. Specifically, 57 papers release data, 15 offer model weights, and 65 share code. Most open-source papers release data and code, while only a tiny fraction provide model weights.  This trend may be attributed to the relative simplicity of earlier models, which reduced the need for sharing model weights. In contrast, publicly available code and data are traditionally in higher demand. However, the emergence of LLMs increases the necessity of providing model weights due to the high cost of reproducing these models. Consequently, this trend may shift, leading to a higher prevalence of open model weights.

Future models are becoming increasingly complex regarding the context they process and their underlying architectures. We advocate adopting open-source practices, encouraging researchers to make their data, code, and model weights publicly available. These practices can enhance the credibility of research papers and significantly support continuous innovation.

\begin{center}
    \begin{myboxc}{\textbf{\ding{43} RQ4 Summary: }
    \begin{enumerate}
        \item Code intelligence tasks cover a wide range of evaluation benchmarks and metrics. However, current methods evaluate predominantly end-to-end performance, lacking sufficient focus on assessing context utilization. Notably, seven papers still rely on human evaluation, suggesting more automated and standardized metrics are needed to reduce subjectivity and improve scalability.
        \item We observe an uneven use of context across different code intelligence tasks, often linked to gaps in datasets that lack certain types of context. Many forms of context remain underutilized, presenting opportunities for future research to expand context coverage in datasets and tasks.
        
        \item We summarize the open-source status of the reviewed papers, highlighting the importance of releasing code, datasets, and model weights. We encourage researchers to make these resources available to foster reproducibility and further advancements in the field.
    \end{enumerate}
    }
    \end{myboxc}
\end{center}

\section{Case Study} \label{case_study}

\begin{sidewaystable}[]
    \vspace*{15cm}
    \renewcommand\arraystretch{1.8}
    \setlength{\LTpre}{0pt}
    \setlength{\LTpost}{0pt}
    \belowrulesep=0pt
    \aboverulesep=0pt
    \setlength{\tabcolsep}{2pt}
    \centering
    \tiny
\caption{Performance Comparison and Context Contribution Analysis Across Code Intelligence Tasks}
\label{tab:performance_comparison}
\begin{tabular}{|c|c|c|c|c|c|c|c|c|c|c|c|}
\toprule
\textbf{Task} & \textbf{Context} & \textbf{Paper} & \textbf{Model} & \textbf{Metric} & \textbf{Benchmark}  & \textbf{Improved Method} & \textbf{Base Performance}  & \textbf{Improved Performance} & \textbf{Relative Improvement}\\
\midrule
Code Generation & Source Code & Liu et al.~\cite{N66}  & CodeT5 & EM & self-created  & NL+Libs+Imports(Gen)+Imports(Ret) & 0.191
& 0.225 & 29.76 \\
\midrule
Code Generation & API documents & Zan et al.~\cite{N4} &CodeGen 350M & Pass@1 & TorchDataEval  & Top 1 API doc & 6.72 & 8.72 & 29.76 \\
\midrule
Code Generation & Compilation Information & Bi et al.~\cite{F4} & GPT-3.5-Turbo & Pass@5 & CodeEval   & CoCoGen & 34.44 & 43.01 & 24.88 \\
\midrule
Code Completion & Source Code & Zhang et al.~\cite{N31} & CodeGen 350M & EM & RepoEval & RepoCoder Iteration 1 & 22.19 & 31.75 & 43.08 \\
\midrule
Code Completion & CFG  & Liu et al.~\cite{F27} & GPT-3.5-Turbo & EM & RepoEval-Updated & GraphCoder & 39.15 & 46.60 & 19.03 \\
\midrule
Code Completion & DFG & Cheng et al.~\cite{F16} & CodeGen 350M & EM & CrossCodeEval & DraCo  & 10.13 & 13.02 & 28.53 \\
\midrule
Code Completion & Compilation Information & Agrawal et al.~\cite{N1} & CodeGen 350M & Compilation Rate & PRAGMATICCODE  & CG-350M-MGD & 52.43 & 65.37 & 24.69 \\
\midrule
Code Completion & IDE & Bibaev et al.~\cite{F30} & Decision Tree & Recall@1 & self-created  & CatBoost & 0.761  & 0.870 & 14.32 \\
\midrule
Code Summarization & Source Code & Guo et al.~\cite{N26}  & LSTM-Attention & METEOR & self-created  & 14.46 & MBT & 16.16 & 11.76 \\
\midrule
Code Summarization & AST & Guo et al.~\cite{N26} & LSTM-Attention & METEOR & self-created & MBT\&AST Analysis & 16.16 & 16.47 & 1.92 \\
\midrule
Code Summarization & API documents &  Shahbazi et al.~\cite{N24} & Transformer & METEOR & CodeSearchNet  & API & 7.60 & 7.82 & 2.89 \\
\midrule
Code Summarization & Code Comments & Zhou et al.~\cite{F52} & CodeLlama 7B & BLEU & mixed project  & doc comments & 22.50 & 41.00 & 82.22 \\
\midrule
Code Summarization & UML &Wang et al.~\cite{N35} & GNN & BLEU & self-created   & CoCoSum & 17.19 & 19.04 & 10.76 \\
\midrule
Commit Message Generation & Source Code & Xu et al.~\cite{N18}& Transformer & BLEU & self-created   & COMEG & 12.12 & 13.47 & 11.14\\
\midrule
Commit Message Generation & AST& Xu et al.~\cite{N18} & Transformer & BLEU & self-created  & COMEG & 12.81  & 13.47 & 5.15 \\
\midrule
Commit Message Generation & Code Diffs& Vu et al.~\cite{N23}& UniXCoder & METEOR & self-created   & context-encoded change & 5.45 & 6.01 & 10.28\\
\midrule
Commit Message Generation & Commit Messages& Eliseeva et al.~\cite{F44} & CodeT5 & B-Norm & CommitChronicle  & CodeT5: History   & 15.12 & 16.8 & 11.11 \\
\midrule
Commit Message Generation & CPG&  Mandli et al.~\cite{F20}& CodeT5 & METEOR & CodeSearchNe  & containing added, deleted, and context token & 11.49 & 12.26 & 6.70 \\
\midrule
Defect Detection & CFG & Zhang et al.~\cite{N49} & CFGNN, GGNN & Precision & self-created   & NodeEmbedding+CFE & 52.30 & 54.20 & 3.63 \\
\midrule
Program Repair & Source Code& Li et al.~\cite{N2} & Transformer & Acr@1 & DeepFix   & TransRepair & 45.67 & 49.65 & 8.71\\
\midrule
Program Repair & Compilation Information& Li et al.~\cite{N2} & Transformer & Acr@1 & DeepFix  & TransRepair & 47.09  & 49.65 & 5.44 \\
\midrule
Program Repair & Code Diffs& Lin et al.~\cite{U02}& LSTM & F1 & self-created   & CACHE & 68.30 & 78.00 & 12.40\\
\midrule
Program Repair & AST & Lin et al.~\cite{U02}& LSTM & F1 & self-created   & CACHE & 71.40& 78.00 & 8.20 \\
\bottomrule
\end{tabular}
\end{sidewaystable}

We compiled representative ablation studies that measure how different context types affect code-intelligence models. Table~\ref{tab:performance_comparison} summarizes the measured performance gains when models are provided with auxiliary context.

Each row reports a single experiment; the table’s primary focus is the relative improvement from adding context. For reference, each row also lists the task, benchmark, model, metric, context type, and the method used to obtain the gain.

The table confirms that incorporating context is a critical enhancer across all code intelligence tasks, consistently yielding non-trivial performance improvements. The observed relative improvements, ranging significantly across tasks and contexts, reveal several key insights regarding context efficacy:

\begin{itemize}
    \item \textbf{High-Level Semantics and Documentation are Paramount.} The most substantial relative improvements are often seen when leveraging external, high-level, or human-curated contexts. For example, incorporating code comments resulted in an 82.22\% improvement in code summarization~\cite{F52}, and using API documents led to a 29.76\% gain in code generation~\cite{N4}. This demonstrates that domain knowledge beyond the immediate source code can yield the most dramatic increases.

    \item \textbf{Graph-based Context is Important for Prediction.} For prediction-based tasks like Code Completion, contexts capturing program execution and data flow are crucial. The inclusion of DFG (28.53\% improvement~\cite{F16}) and CFG (19.03\% improvement~\cite{F27}) validates the need to model operational dependencies to achieve state-of-the-art performance explicitly.

    \item \textbf{Compilation Information Context Maintains Consistency.} Contexts related to the development environment, such as compilation information (e.g., 24.69\% for code completion~\cite{N1}), prove vital for ensuring the practical utility and correctness of the generated or completed code.

\end{itemize}

This data strongly supports the recommendation that future code intelligence systems move beyond simple source code analysis and prioritize the integration of diverse contexts to maximize performance gains across tasks.
\section{Related work} \label{sec:relatedwork}

The burgeoning field of code intelligence has attracted substantial research attention~\cite{All2018ML,Wan2024DL,zheng2024LLM}. While existing surveys predominantly concentrate on model architecture analysis, this study pioneers the first systematic review of context utilization in code intelligence, a critical yet underexplored perspective.

Prior studies in other domains interpret context through domain-specific dimensions. In ubiquitous computing, Hoareau et al. ~\cite{Hoareau2009ModelingAP} conceptualize context through multi-dimensional attributes such as environmental states, geospatial position, and social relations. Recommender systems research~\cite{Ding2019contextSurvey, Rodríguez-Hernández2021contextSurvey,Villegas2018contextSurvey} adopts user-centric contextual paradigms, emphasizing personal profiles and behavioral patterns. Other domains, such as IoT systems~\cite{Perera2014contextSurvey}, intelligent transportation~\cite{Huang2023contextSurvey}, and mobile computing domains~\cite{Rivero-Rodriguez2016contextSurvey}, also investigate the concept in their area and analyze context-aware models. Compared with these studies, our research is the first systematic review of context utilization in code intelligence tasks, clarifying the definition of context and developing a taxonomy of contextual elements within the code intelligence domain.

In the code intelligence domain, existing reviews predominantly focus on model-centric analyses. Allamanis et al.~\cite{All2018ML} examine machine learning models in the domain of big code, an early concept of code intelligence. They explore the similarities and differences between programming languages and natural languages, providing an in-depth review of code-relevant model designs and their applications. Yang et al.\cite{Yang2022DL} review integrating deep learning techniques with software engineering, examining model selection, data processing methods, and relevant datasets. Wan et al.\cite{Wan2024DL} provide a comprehensive literature review on deep learning in code intelligence, focusing on model performance and benchmarking 18 leading models across five code intelligence tasks. Zheng et al.~\cite{zheng2024LLM} explore the evolution of large language models in software engineering by analyzing 134 studies, evaluating code LLMs across various code intelligence tasks, and outlining future directions for model development. In contrast, our research highlights the critical role of context in code intelligence tasks. We present a comprehensive discussion spanning rule-based, feature-based, deep learning-based, and LLM-based models across seven code intelligence tasks. We offer a more holistic perspective on how contextual information can enhance model performance and effectiveness. Moreover, we collate context-related datasets, analyze relevant metrics, and assess the current state of open-source resources in this area. This approach enables us to identify key opportunities and challenges existing studies may have overlooked, paving the way for further advancements in code intelligence.

\section{Challenges and Opportunities}\label{sec:challenges_and_opportunities}

Our analysis identifies three core challenges in context utilization in current code intelligence  systems: 
(1) due to the numerous potential context combinations and diverse utilization approaches, it remains unclear whether the current strategies are optimal for context usage in code intelligence tasks.
(2) existing methods for context processing are unexplored to determine whether they can generalize across different code intelligence scenarios; 
and (3) current benchmarks and evaluation metrics cannot adequately reflect the effectiveness of context utilization.

These challenges motivate three research opportunities:

\begin{itemize}
    \item \textbf{Opportunity 1: Integrating multiple contexts for code intelligence tasks.}
    Different types of contexts provide distinct benefits in addressing code intelligence tasks. For example, compiler information offers detailed execution data and dynamic runtime insights, while UML diagrams present a high-level architectural view of a repository. However, existing research often focuses on a single context type or a limited combination of contexts, leaving the potential of integrating multiple contexts underexplored. While using multiple contexts may introduce additional noise, it also enriches the information available to the model. This creates both challenges and new research prospects. One promising direction is to design adaptive retrieval mechanisms that automatically adjust the types and scope of context based on task requirements and retrieval conditions, thereby improving task performance while managing computational costs. For instance, DomainCodeBench~\cite{zheng2025generalperformancedomain} focuses on LLM performance across various domains with different contexts, discussing the contribution of source code context retrieved based on similarity, dependency, and API in code generation tasks. GraphCoder~\cite{F27} provides an ablation study to highlight the effects of CFG context, Data Dependency Graph (DDG) context, and Control Dependency Graph (CDG) context (the latter two are components of the PDG). While these studies have begun to offer insight, the space remains vast.

    \item \textbf{Opportunity 2: Developing effective context utilization mechanisms.}
    Handling multiple contexts can increase the complexity of models, but it also enhances their capacity to tackle more sophisticated tasks, such as repository-level code intelligence. One potential solution is to employ advanced representation learning techniques to capture semantic information more effectively. Although researchers have explored representation learning methods like COMET~\cite{F20} and MGVD~\cite{F3}, and achieved notable performance on related code intelligence tasks. However, the full potential of combining various context representations remains largely underexplored, and these methods often focus on performance improvements, leaving room for further investigation into time costs. Alternatively, recent advances in large language models offer promising opportunities to construct robust RAG strategies for relevant context, such as DocPrompting~\cite{N6}, RLCE~\cite{F9}, and CoCoGen~\cite{F4}. A key challenge is that these frameworks often focus on specialized contexts to solve specific tasks, limiting the validation of a unified pipeline that incorporates diverse context-retrieval strategies across multiple code intelligence tasks. Moreover, future work should move beyond offline evaluation and systematically explore the timing and frequency of context extraction (e.g., offline caching and incremental updates) to bridge the gap between research prototypes and real-world production tools.

    \item \textbf{Opportunity 3: Constructing robust evaluations for context-aware models.}
    Current evaluation methods for context-aware models often struggle to adapt when multiple contexts are introduced. Future research should focus on developing new benchmarks and evaluation metrics to address these issues. A comprehensive, multidimensional framework is needed to benchmark various code intelligence tasks and context types. While recent research has made some initial efforts, more work is required. For example, Coderag-bench~\cite{wang2025coderag} provides five document sources for retrieval and is specialized for eight types of coding tasks. For metrics, more fine-grained evaluation schemes should be explored, such as measuring the efficiency and effectiveness of contextual information processing. As exemplified by RepoExec~\cite{F32}, which proposes the ``Dependency Invocation Rate'', a metric specifically designed to quantify how well a model uses contextual information for dependency generation. Crucially, to enable the precise context evaluation, researchers need to provide more detailed ground truth annotations specific to the context when releasing benchmarks. This fine-grained assessment of context-aware metrics is an area that requires further dedicated research.

\end{itemize}
\section{Conclusion and Future Work}
\label{sec:conclusion}

In this paper, we explore the utilization of context in code intelligence. We analyze the research trends in this area from 146 context-relevant studies across seven code intelligence tasks. We categorize different types of context used in these tasks and examine researchers' preferences for context utilization. Our findings indicate that context application in code intelligence is still insufficient or underexplored in several areas. We also review the preprocessing techniques and modeling approaches, identifying opportunities for optimizing current methods, especially with the advancements in large language models (LLMs). Additionally, we assess metrics, datasets, and open-source status in these studies.

Our study offers a comprehensive overview of context utilization in code intelligence, highlighting potential opportunities for future research. (1) Future work should focus on better leveraging different types of context to improve code intelligence tasks. (2) There is also a need to enhance the efficiency and performance of context utilization with the advancements in large language models. (3) Developing more effective benchmarks and metrics to evaluate context utilization is essential. These areas represent promising directions for continued exploration and development.
\section{Acknowledgments}
This work is partially supported by CCF-Huawei Populus Grove Fund CCF-HuaweiSE202403.

\bibliographystyle{ACM-Reference-Format}
\bibliography{ref}

\end{document}